\documentclass[aps,twocolumn,secnumarabic,balancelastpage,amsmath,amssymb,nofootinbib,floatfix,superscriptaddress]{revtex4-2}
\usepackage{graphicx}      
\usepackage{siunitx}
\usepackage{xcolor}
\usepackage{braket}
\usepackage{float}
\usepackage{booktabs,caption}
\usepackage{array}
\usepackage[flushleft]{threeparttable}
\usepackage{textcomp}
\usepackage{adjustbox}
\usepackage{tabularx}
\usepackage{xcolor,colortbl}
\definecolor{Gray}{gray}{0.85}
\definecolor{LightCyan}{rgb}{0.88,1,1}
\usepackage[colorlinks=true]{hyperref}  

\newcommand{\modif}[1]{\textcolor{black}{#1}}

\begin{document}
\raggedbottom

\title{Easy-plane ferromagnetic ordering and crystal-field ground state in the Kondo lattice CeCuSi}
\author{Hanshang Jin}
\affiliation{Department of Physics and Astronomy, University of California, Davis, California 95616, USA}
\author{Owen Moulding}
\affiliation{Univ. Grenoble Alpes, CNRS, Grenoble INP, Institut Néel, 38000 Grenoble, France}

\author{James C. Fettinger}
\affiliation{Department of Chemistry, University of California, Davis, California 95616, USA}
\author{Yingzheng Gao}
\affiliation{Univ. Grenoble Alpes, CNRS, Grenoble INP, Institut Néel, 38000 Grenoble, France}
\author{Peter Klavins}
\affiliation{Department of Physics and Astronomy, University of California, Davis, California 95616, U.S.A.}
\author{Marie-Aude Méasson}
\affiliation{Univ. Grenoble Alpes, CNRS, Grenoble INP, Institut Néel, 38000 Grenoble, France}
\author{Valentin Taufour}
\affiliation{Department of Physics and Astronomy, University of California, Davis, California 95616, U.S.A.}

\begin{abstract}
We report the successful growth of CeCuSi single crystals using a metallic flux method and the physical properties using structural, magnetic, electrical transport, optical, and heat capacity measurements. CeCuSi crystallizes in a hexagonal-bar shape, and single crystal x-ray diffraction confirms the ZrBeSi-type structure (space group $P6_{3}/mmc$). CeCuSi orders ferromagnetically below $T_\textrm{C}=15.5$\,K with easy magnetization direction within the basal plane. The Ce$^{3+}$ ions are situated within a triangular lattice with a point group of $D_{3d}$. We perform a detailed crystalline electric field (CEF) analysis of the anisotropic magnetic susceptibility, the Schottky anomaly in heat capacity, and the Raman-active excitations. The results indicate a ground state doublet with magnetic moment primarily in the basal plane, and a ferromagnetic interaction along both directions. The exponential behavior in resistivity and in heat capacity below $T_\textrm{C}$ can also be well explained by the ferromagnetic magnon model. We found that CeCuSi does not exhibit the CEF hard axis ordering observed in many ferromagnetic Kondo lattice compounds. Our CEF analysis suggests that the exchange interactions along both axes are ferromagnetic, potentially
explaining the absence of hard-axis ordering.

\end{abstract}

\maketitle


\section{Introduction}
$RTX$ ($R$ = rare-earth, $T$ = $d$-electron transition-metal, and $X$ = $p$-electron elements) series is a large family of intermetallic compounds that exhibit a wide range of interesting physical properties such as heavy fermion behavior, Kondo effect, electride catalyst, superconductivity, and a variety of magnetic orderings~\cite{ye2017Copperbased,gupta2015Review}. In particular, those of cerium (Ce$TX$) have attracted broad interests with respect to their magnetic properties and two different cerium valence states. The CeTX family demonstrates a wide variety of interesting phenomena such as valence fluctuation, magnetic orderings, superconductivity, or non-Fermi liquid systems, depending on the valence state of the Ce ion~\cite{pottgen2015Cerium, pottgen2015Equiatomic, pottgen2016Cerium, janka2016Cerium}. Among them, CeCuSi is one of the rare Ce compounds that exhibits a ferromagnetic (FM) ordering with a Curie temperature of $T_\textrm{C} = 15.5$\,K, the highest in the Ce$TX$ family.

More interestingly, the Ce$^{3+}$ ions are situated within a triangular lattice with a point group of $D_{3d}$ that is geometrically frustrated.
Geometrically frustrated magnetic systems have attracted significant interest due to the exotic ground states, such as spin liquid~\cite{balents2010Spin}, spin ice~\cite{huang2014Quantum,sibille2015Candidate},  and spin glass~\cite{kirkpatrick1977Frustration}.
While the majority of studies on triangular lattices focus on antiferromagnetic compounds, where geometrical frustration is evident ~\cite{sibille2015Candidate,uzoh2023Influence,ochiai2021Fieldinduced}, 
here we report on a ferromagnetically ordered Ce compound with a triangular lattice.

So far, studies of CeCuSi were limited to polycrystalline samples. In previous studies with polycrystalline samples, the structure was reported to be either a disordered AlB$_2$-type ($P6/mmm$, 191) structure~\cite{rieger1969Ternare}, or an ordered phase. For the former, the disorder comes from the interchange of Cu and Si atoms, whereas for the latter, either ZrBeSi-type or Ni$_2$In-type ($P6_{3}/mmc$, 194) structures were present~\cite{gignoux1986Magnetic,iandelli1983Low,sondezi-mhlungu2009Crystal}, or a mixture of both ordered and disordered secondary CeCu$_{0.8}$Si$_{1.2}$ phases~\cite{yang1991Magnetic}. The neutron diffraction on the polycrystalline sample suggests a collinear FM ordering along an easy magnetization plane, with a magnetic moment of 1.25\,$\mu_\mathrm{B}$~\cite{gignoux1986Magnetic}.
Magnetization measurements under pressure showed an increase of the Curie temperature reaching 30\,K at 10\,GPa suggesting an increased hybridization of Ce-$4f$ with Ce-$5d$ and Cu-$3d$ electrons~\cite{hearne2014Pressure}. The decrease of the magnetization combined with XANES and its dichroism (XMCD) at the Ce $L_3$-edge indicate a pressure induced delocalization of the $4f$ electron~\cite{hearne2014Pressure}.

In this study, we successfully grew single crystals of CeCuSi using a self-flux solution growth. We confirm the ZrBeSi-type structure with a $P6_{3}/mmc$ space group~\cite{iandelli1983Low}. We observe the exponential magnon behavior at low temperatures both in resistivity and heat capacity measurements. We also perform a detailed crystalline electric field (CEF) analysis on the magnetic susceptibility data along both axes and the heat capacity data. The curvature behavior in inverse magnetic susceptibility data and the Schottky anomaly in heat capacity measurement can be well explained by the CEF model. Similarly, with polarized Raman spectroscopy, we observe CEF excitations that are consistent with the CEF model. The results from the CEF analysis indicate that the $ab$ plane serves as the easy magnetization direction. However, many of the Ce- and Yb-based Kondo lattice (KL) ferromagnets order along the magnetically CEF hard direction~\cite{hafner2019Kondolattice}. Our CEF analysis suggests that the interactions parallel and perpendicular to the $c$-axis are ferromagnetic, potentially
explaining the absence of hard-axis ordering.


\section{Experimental Procedure}
\begin{figure}[!htb]
\center
\includegraphics[width=\linewidth]{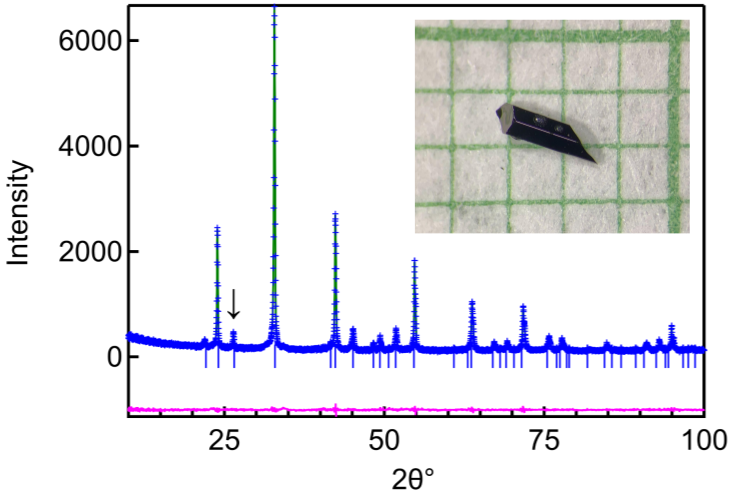}
\caption{X-ray powder diffraction pattern of CeCuSi with the inset photo of the single crystal on a millimeter grid. The arrowed peak is unique to the space group $P6_{3}/mmc$\,(194) compared to $P6/mmm$\,(191).}
\label{fig:crystal_xrd}
\end{figure}

Single crystals of CeCuSi were synthesized via the self-flux method. In order to identify a flux composition, we first prepared several mixtures of Ce-Cu-Si using an arc-melter, and identified the crystallographic phases using powder x-ray diffraction. Further optimization of the temperature profile led to the following procedure. The materials  [Ce pieces (Ames Lab), Cu pieces (5N), Si pieces (7N)], with a starting composition of Ce$_{42}$Cu$_{42}$Si$_{16}$, were arc-melted and sealed in a tantalum crucible with a tantalum strainer~\cite{jesche2014Single}, then sealed in a fused silica ampoule in a partial pressure of argon. The sealed ampoule was placed in a furnace where it was held at \SI{1150}{\celsius} for 10 h, slowly cooled to \SI{750}{\celsius} in two weeks, and then the ampoule was removed from the furnace and quickly centrifuged to separate the single crystals from the molten flux. The crystals formed hexagonal bar shapes with shining surfaces, as shown in the inset of Fig.~\ref{fig:crystal_xrd}. We noticed that 1-2 days after the synthesis, some crystals started to form cracks and break into smaller pieces on their own. However, this does not seem to be due to the sample decomposing into different structures or different phases, nor that the samples are air/moisture sensitive under ambient conditions. After a few days, the samples no longer formed cracks and remained as one solid piece with a shiny surface. The PXRD pattern has no significant difference after ten days, which rules out a structural change. The single-crystal x-ray diffraction was performed after this period and confirms that the structure remains in the hexagonal $P6_{3}/mmc$ ordered structure. We suspect that the cracks in the samples are due to defects formed during the solution growth process\modif{, or due to the thermal stress during cooling after centrifuging at \SI{750}{\celsius}}. Remaining stress and/or defects in the single crystals will be evident in the electrical resistivity and Raman spectroscopy measurements presented in this paper.

Single-crystal x-ray diffraction (SCXRD) data were collected at 90\,K using a sealed-tube Mo x-ray source on a Bruker APEX-II CCD. The refinement results, confirming that our single crystals adopt an ordered $P6_{3}/mmc$ structure, will be discussed in the next section. 

The phase identification of the samples was also carried out by powder x-ray diffraction (PXRD) on a Rigaku Miniflex 600 diffractometer with Cu K$\alpha$ ($\lambda$ = 1.54178\,\si{\angstrom}) radiation at room temperature as shown in Fig.~\ref{fig:crystal_xrd}. The PXRD pattern collected from crushed single crystals shows a clear peak at $2\theta = 26.42^{\circ}$, which is unique to the space group 194 compared to the space group 191, further confirming that our single crystals have the hexagonal $P6_{3}/mmc$ (194) ordered structure. Rietveld refinements of the PXRD pattern were performed using the program GSAS II~\cite{toby2013GSASII}. The lattice parameters $a = 4.238\,\si{\angstrom}$, $c = 7.990\,\si{\angstrom}$, and unit cell volume $V = 124.306\,\si{\angstrom^3}$ were then obtained, in good agreement with the previous reports~\cite{gignoux1986Magnetic,iandelli1983Low,mugnoli1984Neutron,sondezi-mhlungu2009Crystal} and our SCXRD results. 

A polycrystalline sample of LaCuSi was prepared by arc-melting the constituents under an argon atmosphere. The alloyed-pellet was flipped over four times on remelting for proper homogenization. It was annealed at \SI{800}{\celsius} for two weeks. Similarly, the PXRD of our polycrystalline LaCuSi also shows the unique peak from the space group 194.

Magnetic properties were measured using a SQUID magnetometer (Quantum Design MPMS XL) with 7\,T magnetic field, in the temperature range of 2$-$300\,K. The bar-shaped sample was sandwiched between two cocktail straws when measuring along the $c$-axis, to ensure it was well centered radially~\cite{ullah2024Experimental}. For measurements with the field in the $ab$ plane, the sample was sandwiched in the gap between two cut straws. This method avoids the use of grease, which reduces the background susceptibility ($\chi_0$) in the measurements, thereby allowing for a more accurate CEF analysis.

Resistivity measurements were carried out in a Quantum Design Physical Property Measurement System (PPMS) from 1.8$-$300\,K. The resistivity was measured by the standard four-probe method using Pt wires (0.001 inch diameter) applied with silver-filled epoxy (EPO-TEK\textsuperscript{®} H20E) using the ac method ($f=17$\,Hz, $I = 1$\,mA). 

The PPMS was also used to obtain heat capacity data using the relaxation technique down to 1.8\,K. Due to the small mass of the single crystals (ranging from about 3.5\,mg to 7.5\,mg), the thermal coupling falls at high temperatures. The polycrystalline pellet of 22.4\,mg was prepared by directly grinding selected single crystals from the same batch into a powder and pressing into a pellet without any other operation. This allowed us to maintain a thermal coupling near 99.5\% throughout the entire temperature range, providing a more accurate result above the transition temperature. The presented heat capacity data combines single-crystal data below 20 K, preserving the sharp feature at the transition, with polycrystalline data above 20 K, providing more accurate data in the region of the Schottky anomaly.

The polarized Raman measurements used a 532\,nm solid-state laser with an incident laser power of either 1 or 5\,\si{\milli\watt}. Measurements were performed at room temperature and 2\,K base temperature. A Trivista 777 spectrometer equipped with ultra-low-noise, cryogenically cooled PyLon CCDs was used in triple-stage subtractive mode. The sample was orientated such that the incident laser light was perpendicular to the $ac$ plane.
\section{Results and Discussion}

\subsection{Crystal Structure}
The single-crystal x-ray structure refinement results are shown in Table~\ref{cs_refine}, and the refined atomic positional and displacement parameters are shown in Table~\ref{cs_refine_pos}. The refined lattice parameters $a = 4.239\,\si{\angstrom}$, $c = 7.972\,\si{\angstrom}$, and unit cell volume $V = 124.083\,\si{\angstrom^3}$ were obtained, and the values are very close to the refined values from PXRD mentioned previously.
The final refinement residual is below 2\% and the results show no evidence of mixed occupancy between Cu and Si, with each site refining to 100\% of the initial composition. If any mixed occupancy is present, it is likely below 1\%, which is undetectable using SCXRD data.

\begin{table}[!htb]
\centering
\caption{Crystal structure refinement results of CeCuSi}
\label{cs_refine}
\resizebox{\columnwidth}{!}{
\begin{tabular}{lc}
\hline
Refined items          		& Values                        \\ \hline
Chemical formula         & CeCuSi    \\
Chemical formula weight  	& 231.75 $g/mol$    \\
Temperature         & 90(2) K    \\
Wavelength      & 0.71073 \si{\angstrom}   \\
Crystal system        	& Hexagonal     \\
Space group          	& $P6_{3}/mmc$     \\
Unit Cell dimensions 	& $a$ = 4.23940(14)~\si{\angstrom}, $\alpha$ = 90$^{\circ}$   \\
	& $b$ = 4.23940(14)~\si{\angstrom},	$\beta$ = 90$^{\circ}$ \\
 	& $c$ = 7.9721(4)~\si{\angstrom}, 	$\gamma$ = 120$^{\circ}$\\
Cell volume & 124.083(10)$~\si{\angstrom}^3$ \\
$Z$ & 2 \\
Density (calculated)  & 6.203 g/cm$^3$\\
Absorption coefficient    	& 26.740 mm$^{-1}$\\
F(000)			& 202\\
Crystal size           	& $0.132 \times 0.063 \times 0.045$ mm$^{3}$\\
Crystal color and shape  	& Silver Platelet\\
Diffractometer  	& Bruker APEX-II CCD\\
Theta range    & 5.115 to 30.195$^{\circ}$ \\
Index ranges    & $-6\leq h\leq6$ \\
 & $-6\leq k\leq6$ \\
 & $-11\leq l\leq 11$ \\
Reflections collected  	& 1312\\
Independent reflections  	& 94 [R(int) = 0.0411]\\
Observed reflections (I $>$ 2$\sigma$(I))  	& 86\\
Completeness to $\theta$ = 25.242$^{\circ}$ 	& 100\% \\
Absorption correction	& Semi-empirical from equivalents\\
Max. and min. transmission  	& 0.2451 and 0.1016\\
Solution method   	& SHELXT (Sheldrick, 2014)\\
Refinement method   & SHELXL-2018/3 (Sheldrick, 2018) \\
 & Full-matrix least-squares on F2 \\
Data / restraints / parameters 	& 94 / 0 / 8 \\
Goodness-of-fit on F$^{2}$      	& 1.168 \\
Final R indices [I $>$ 2$\sigma$(I)]     & R$_1$ = 0.0183, wR$_2$ = 0.0467 \\
R indices (all data) 	& R$_1$= 0.0195, wR$_2$ = 0.0474 \\
Extinction coefficient  	& 1.74(13)\\
Largest diff. peak and hole & $1.341/-1.442$\,(e.$\si{\angstrom}^3$)\\ \hline
\end{tabular}}
\end{table}

\begin{table}[!htb]
\centering
\begin{threeparttable}
	\centering
\caption{Atomic coordinates and equivalent ($\times10^4$)   isotropic displacement parameters ($\si{\angstrom}^2\times10^3$) for CeCuSi.  $U_{eq}$ is defined as one-third of the trace of the orthogonalized $U^{ij}$ tensor.
}
	\label{cs_refine_pos}
	\begin{tabularx}{\columnwidth}{>{\arraybackslash}X  >{\arraybackslash}X  >{\arraybackslash}X >{\arraybackslash}X  >{\arraybackslash}X  >{\arraybackslash}X}
	\hline
		&\modif{Wyckoff} & x	 & y	 & z	 & $U_{eq}$ \\ \hline
	Ce1 &\modif{2a} & 0 & 0 & 0 & 6(1) \\ 
	Cu1 &\modif{2c} & 6667 & 3333 & 2500 & 8(1) \\ 
	Si1 &\modif{2d} & 3333 & -3333 & 2500 & 7(1) \\ \hline
	\end{tabularx}

\bigskip

\centering
\begin{tablenotes}
\item
Anisotropic displacement parameters  ($\si{\angstrom}^2\times10^3$) for CeCuSi.

\end{tablenotes}
\begin{tabularx}{\columnwidth}{>{$\arraybackslash$}X  >{$\arraybackslash$}X  >{$\arraybackslash$}X >{$\arraybackslash$}X  >{$\arraybackslash$}X  >{$\arraybackslash$}X >{$\arraybackslash$}X}
\hline
Atom & $U^{11}$ & $U^{22}$ & $U^{33}$ & $U^{23}$ & $U^{13}$ & $U^{12}$ \\ \hline
Ce1 & 8(1) & 8(1) & 4(1) & 0 & 0 & 4(1) \\ 
Cu1 & 8(1) & 8(1) & 7(1) & 0 & 0 & 4(1) \\
Si1 & 9(1) & 9(1) & 2(2) & 0 & 0 & 4(1) \\ \hline
\end{tabularx}
\end{threeparttable}
\end{table}

\subsection{Magnetic properties}
\begin{figure}[!htb]
\center
\includegraphics[width=\linewidth]{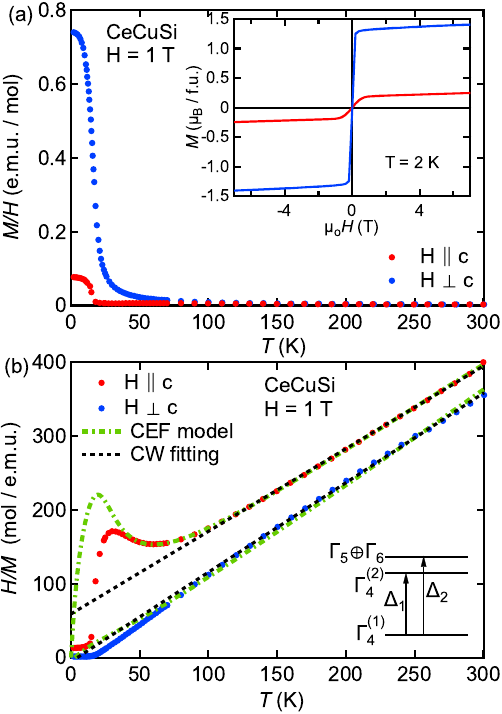}
\caption{(a) Magnetic susceptibility vs temperature of
CeCuSi along the $c$ axis and perpendicular to $c$ axis ($ab$ plane). The inset is the field-dependent magnetization along both axes.
(b) Inverse susceptibility for both axes were fitted to Curie-Weiss law and the CEF model, and inset shows the proposed CEF scheme with $\Delta_1 = 112$\,K and $\Delta_2 = 122$\,K.}
\label{fig:MvsTH}
\end{figure}

Magnetic susceptibility as a function of temperature with an applied field of 1\,T along the $c$ axis and the $ab$ plane from 2$-$300\,K are shown in Fig.~\ref{fig:MvsTH}(a). The high-temperature susceptibility data along both axes from 100$-$300\,K were fit to the Curie-Weiss law, as shown in Fig.~\ref{fig:MvsTH}(b).

The effective paramagnetic moments $\mu_{\mathrm{\scriptscriptstyle{eff}}}$ are $2.67$\,$\mu_\mathrm{B}$ and $2.56$\,$\mu_\mathrm{B}$ along the $c$ axis and $ab$ plane respectively, which is close to the theoretical Ce$^{3+}$ value ($2.54$\,$\mu_\mathrm{B}$). The Curie-Weiss temperatures along the $c$ axis and $ab$ plane are $\theta _{CW}^{\parallel} = -52.6$\,K and $\theta _{CW}^{\perp} = 5.67$\,K, respectively. The significant difference and negative Curie-Weiss temperatures are not attributed to an antiferromagnetic interaction, but rather to the CEF effect. The detailed analysis of the CEF effect is presented in the next section.

The isothermal magnetization curves along the $c$ axis and $ab$ plane are shown in the inset of Fig.~\ref{fig:MvsTH}(a). The $ab$ plane is the easy magnetization plane with a quick saturation moment of 1.25\,$\mu_\mathrm{B}$ at 0.2\,T, and a smaller saturation moment of 0.18\,$\mu_\mathrm{B}$ at 0.9\,T along the $c$ axis. The small ferromagnetic component in the $c$-direction is not due to a misalignment of the sample, although the sample does indeed experience strong torque to rotate back to the $ab$ plane. However, this can be well explained by our CEF model.
\modif{The isothermal magnetization curves shown in the inset of Fig.~\ref{fig:MvsTH}(a) include both field-raising and field-lowering measurements and the data overlap with each other.} The hysteresis loops are almost invisible along both axes indicating that single crystals lack defects able to pin the magnetic domain walls. This provides some evidence of the relatively good quality of the samples, despite the sample cracking noticed earlier, and the relatively high residual resistivity that will be discussed later.

\subsection{Crystalline Electric Field from Magnetic Susceptibility}
CEF analysis in a Ce-based system is crucial because it prevents misinterpretations of magnetic properties arising from the CEF effects, which can otherwise lead to incorrect conclusions about the nature of exchange interactions based on Curie-Weiss temperatures fits of susceptibility data~\cite{ajeesh2017Isingtype}. In CeCuSi, the Ce atoms have a point group of $D_{3d}$, and the corresponding theoretical magnetic susceptibility expressions were obtained by solving the CEF Hamiltonian. The details can be found in the Appendix~\ref{Tri_CEF}.

The inverse magnetic susceptibility is fit to the experimental data for both orientations simultaneously, and the fitted curves are shown in Fig.~\ref{fig:MvsTH}(b). The CEF parameters are listed in Table~\ref{fittings}. The corresponding ground state is $\Gamma_{4}^{(1)} = 0.257\ket{\pm 5/2} - 0.967\ket{\mp 1/2}$, the first excited state is $\Gamma_{4}^{(2)} = -0.967\ket{\pm 5/2} - 0.257\ket{\mp 1/2}$ with a splitting energy of $\Delta_1 = 112$\,K, and the second excited state is $\Gamma_{5}\oplus\Gamma_{6}=\ket{\pm 3/2}$ with $\Delta_2 = 122$\,K. \modif{Our CEF model effectively explains the curvature behavior in the inverse magnetic susceptibility. However, it does not account for magnetic ordering or for the Kondo effect, which may contribute to the deviation observed at low temperatures.}

The expected saturation magnetization parallel and perpendicular to the $c$ axis from the ground state are given by:

\begin{equation}
\begin{split}
M^{sat}_{\parallel c} & = \Braket{\Gamma_{4}^{(1)}|J_z|\Gamma_{4}^{(1)}} g_{J}\mu_\mathrm{B} = 0.26\,\mu_\mathrm{B}/\mathrm{Ce}^{3+},\\
M^{sat}_{\perp c} & = \Braket{\Gamma_{4}^{(1)}|J_x|\Gamma_{4}^{(1)}} g_{J}\mu_\mathrm{B} = 1.20\,\mu_\mathrm{B}/\mathrm{Ce}^{3+}.\\
\end{split}
\label{eqTi}
\end{equation}
These calculated values agree well with our experimental data, where we see a quick saturation with a moment of 0.18\,$\mu_\mathrm{B}$ along the $c$ axis, and a moment of 1.25\,$\mu_\mathrm{B}$ along the $ab$ plane.

\begin{table}[!htb]
\caption{Fitting parameters to the inverse magnetic susceptibility data based on the Eq.~(\ref{eq:inverchi}):}
\begin{tabular}{p{3cm}p{5cm}}\hline
$B_{2}^{0} = 4.42$\,K & $\chi_{0}^{\parallel}=5.63\times10^{-5}$ (e.m.u./mol)     \\
$B_{4}^{0} = -0.293$\,K & $\chi_{0}^{\perp}=-8.21\times10^{-5}$ (e.m.u./mol) \\
$B_{4}^{3} = 2.93$\,K & $\lambda_{\parallel}=9.0$ (mol/e.m.u.)    \\
    & $\lambda_{\perp}=5.2$ (mol/e.m.u.) \\\hline
\end{tabular}
\label{fittings}
\end{table}

The previous inelastic neutron scattering data~\cite{sondezi-mhlungu2009Crystal} at $100$\,K found that $B_{2}^{0} = 2.43$\,K, $B_{4}^{0} = -0.28$\,K, and $B_{4}^{3} = 5.43$\,K. Thus the derived ground state is $\Gamma_{4}^{(1)} = 0.497\ket{\pm 5/2} - 0.868\ket{\mp 1/2}$, the first excited state is $\Gamma_{4}^{(2)} = -0.868\ket{\pm 5/2} - 0.497\ket{\mp 1/2}$ with a splitting energy of $\Delta_1 = 119$\,K, and the second excited state is $\ket{\pm 3/2}$ with $\Delta_2 = 128$\,K. These previous results from polycrystalline samples agree with our single crystal data and CEF analysis.

A list of Ce compounds with a trigonal site symmetry and a detailed CEF analysis is shown in Table~\ref{trigonal_Ce_list}. The majority of the compounds studied are either AFM or show no ordering at low temperatures and are quantum spin ice or quantum spin liquid candidates. By comparison, CeCuSi has the lowest CEF splitting energy and the highest ordering temperature.

\begin{table*}[!htp]
\caption{List of Ce compounds with a trigonal site symmetry that have a detailed CEF analysis. $*$ = compounds that have a CEF analysis with CEF parameters $B_{6}^{n}$, assuming the $J=7/2$ states are accessible.}
\begin{tabular}{|p{2.8cm}|p{2.2cm}|p{1cm}|p{3cm}|l|l|l|p{1cm}|p{1cm}|p{1.6cm}|}\hline
Compound    & Ordering & Ce point group & Ground State & $B_{2}^{0}$ (K) & $B_{4}^{0}$ (K) & $B_{4}^{3}$ (K) & $\Delta_1$\,(K)    & $\Delta_2$\,(K)  & Mixing Angle $\alpha$ in the ground state \\ \hline

CeCuSi [This work]  & $T_\textrm{C} = 15.5$\,K & $D_{3d}$ & $\Gamma_{4}^{(1)}$, favor $ab$ plane & 4.42 & -0.293& 2.93& 112  & 122 & $-75.1^{\circ}$ \\
CeAuSn~\cite{huang2015Lowtemperature}   & $T_\textrm{N} = 4.4$\,K & $D_{3d}$ & $\Gamma_{4}^{(1)}$, favor $ab$ plane & 11.0 & -0.58& 19.64 & 344  & 445 & $-61.6^{\circ}$ \\
CeIr$_{3}$Ge$_{7}$~\cite{banda2018Crystalline,rai2018Mathrm}   & $T_\textrm{N} = 0.63$\,K & $S_{6}$ & $\Gamma_{4}^{(1)}$, favor $ab$ plane & 34.4 & 0.82 & 67.3 & 374  & 1398 & $-57^{\circ}$ \\
CeCd$_{3}$As$_{3}$~\cite{uzoh2023Influence, dunsiger2020Longrange}   & $T_\textrm{N} = 0.42$\,K & $D_{3d}$ & $\Gamma_{4}^{(1)}$, favor $ab$ plane & 18.55 & -0.08 & 23.02 & 242  & 553 & $-63.9^{\circ}$ \\
CeCd$_{3}$P$_{3}$~\cite{uzoh2023Influence, higuchi2016Optical}   & $T_\textrm{N} = 0.42$\,K & $D_{3d}$ & $\Gamma_{4}^{(1)}$, favor $ab$ plane & 20.9 & -0.03 & 26 & 257  & 621 & $-63.7^{\circ}$ \\
CeZn$_{3}$P$_{3}$~\cite{ochiai2021Fieldinduced}   & $T_\textrm{N} = 0.8$\,K & $D_{3d}$ & $\Gamma_{4}^{(1)}$, favor $ab$ plane & 7.78 & -0.784 & 16.5 &   364.7 & 370.7 & $-63.7^{\circ}$ \\
CePtAl$_{4}$Ge$_{2}$~\cite{shin2020Magnetic}   & $T_\textrm{N} = 2.3$\,K & $D_{3d}$ & $\Gamma_{4}^{(1)}$, favor $ab$ plane & 13.26 & -0.302 & $\approx$0 & 170  & 257 & N/A \\\hline

Ce$_{2}$Zr$_{2}$O$_{7}$~\cite{gao2019Experimental, gaudet2019Quantum}   & no \,ordering down to 0.06\,K & $D_{3d}$ & $\Gamma_{5}\oplus\Gamma_{6}$ & -14.7 & 3.7 & 21.6 & 639  & 1276 & N/A \\
Ce$_{2}$Hf$_{2}$O$_{7}$*~\cite{poree2022Crystalfield}   & no \,ordering down to 0.08\,K & $D_{3d}$ & $\Gamma_{5}\oplus\Gamma_{6}$ & -9.5 & 2.59 & 20.56 & 650  & 1299 & N/A \\
Ce$_{2}$Sn$_{2}$O$_{7}$*~\cite{sibille2015Candidate}   & no \,ordering down to 0.02\,K & $D_{3d}$ & $\Gamma_{5}\oplus\Gamma_{6}$ & - & - & - & 580  & 870 & N/A \\ \hline

\end{tabular}

\label{trigonal_Ce_list}
\end{table*}  

The positive molecular field contributions in both axes (see Table~\ref{fittings}) are consistent with the ferromagnetic ordering of the compound. The negative Curie-Weiss temperature along the $c$ axis is mainly due to the strong CEF effect, which leads to anisotropic behavior, rather than being explained as an antiferromagnetic interaction. The small negative residual susceptibility in the $ab$ plane ($\chi_{0}^{\perp}$) could be due to the gap created when using straw sandwiching mounting methods while the bar-shaped sample is placed horizontally.

Unlike many Kondo lattice ferromagnets that exhibit peculiar magnetic ordering along the CEF hard axis~\cite{hafner2019Kondolattice}, CeCuSi does not display hard-axis ordering. This might be because the molecular field contributions along both directions are positive in CeCuSi, unlike in the hard-axis ordering KL ferromagnets CeAgSb$_2$, where there is a significant in-plane AFM exchange interaction~\cite{hafner2019Kondolattice}.


\subsection{Heat Capacity Measurement}

\begin{figure}[!htb]
\center
\includegraphics[width=\linewidth]{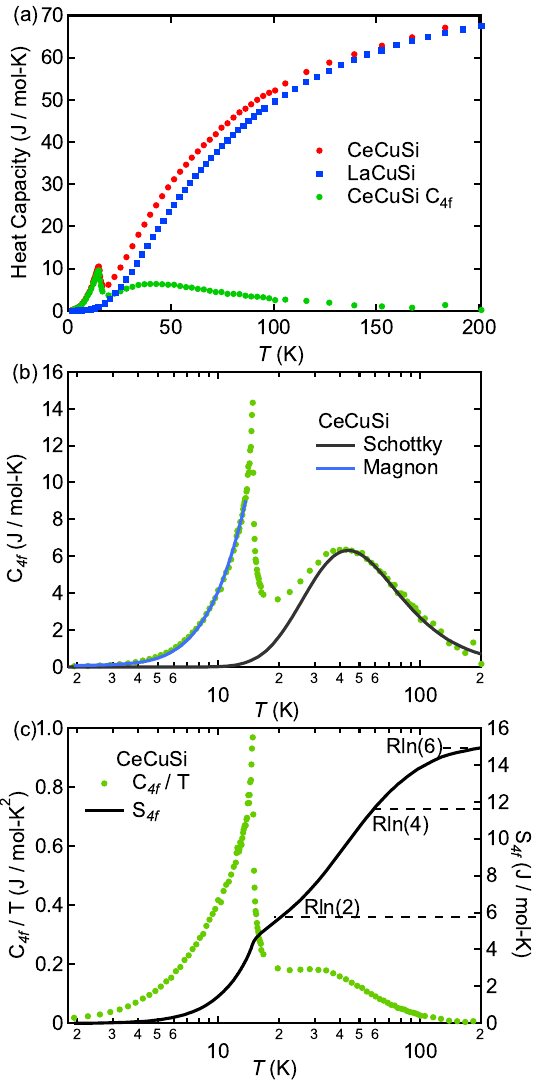}
\caption{(a) The specific heat of single crystal CeCuSi pellet (red), polycrystalline LaCuSi (blue), and their difference (green) from 2$-$200\,K.(b) The $4f$ contribution to the specific heat $C_{4f}$ from 2$-$200\,K in CeCuSi. The blue line shows the fit to ferromagnetic magnon gap expression below $T_\textrm{C}$. The black line shows the fit to Schottky heat capacity due to the CEF. (c) Temperature dependence of $C_{4f}/T$ (left) and of the entropy $S_{4f}$ (right) of CeCuSi. The black dashed lines represent the different expected entropy values $Rln(n)$ due to the CEF doublets.}
\label{fig:HC}
\end{figure}

The specific heat of single crystal CeCuSi pellet, polycrystalline LaCuSi (blue), and their difference (green) that represents the $4f$ contribution to the heat capacity $C_{4f}$ of CeCuSi from 2\,K$-$200\,K are shown in Fig.~\ref{fig:HC}(a).

The magnetic ordering temperature is characterized by the $\lambda$-like anomaly, with a sharp peak at $14.9$\,K. The heat capacity data below $T_\textrm{C}$ can be fitted to the FM spin wave gap expression~\cite{sidorov2003Magnetic,coqblin1977electronic}:
\begin{equation} \label{HCFM}
C_{4f}=\gamma T+\delta (\Delta ^2/\sqrt T+3\Delta \sqrt T+5\sqrt {T^3})e^{-\Delta/T}
\end{equation}
The first term is the usual electronic term and the second term describes the contribution due to an FM spin-wave excitation spectrum with an energy gap $\Delta$. The low-temperature heat capacity was first fit to the linear term $\gamma T$, and it has a good fit up to $3.57$\,K, with $\gamma_{4f} = 41.2$\,\,{mJ mol$^{-1}$ K$^{-2}$}. By fixing the value of $\gamma$, Eq.(\ref{HCFM}) gives a good fit up to 14\,K as shown as the blue line in Fig.~\ref{fig:HC}(b). This yields $\delta = 77.7$\,{mJ mol$^{-1}$ K$^{-5/2}$}, and a magnon gap of $\Delta = 25.3$\,K. Similar exponential gap-like behavior is observed in the electrical resistivity discussed below.

A significant magnon gap, higher than the Curie temperature ($T_\textrm{C}$), often indicates strong magnetic anisotropy. For instance, in the well-studied ferromagnetic compound CeAgSb$_2$, it exhibits $T_\textrm{C} = 9.6$\,K and a magnon gap $\Delta = 24.4$\,K~\cite{jobiliong2005Magnetization}.

The high temperature of the $4f$ contribution in the specific heat measurement can be well characterized by a broad Schottky anomaly resulting from excitations to CEF levels. For a three-level system, the Schottky anomaly can be expressed as follows~\cite{souza2016Specific}:
\begin{equation} \label{HCSCH}
\begin{split}
C_{sch} =\frac{R}{(k_BT)^2}\frac{e^{(\Delta_1+\Delta_2)/k_BT}}{(e^{\Delta_1/k_BT}+e^{\Delta_2/k_BT}+e^{(\Delta_1+\Delta_2)/k_BT})^2}\\
\times[-2\Delta_1\Delta_2+\Delta_2^2(1+e^{\Delta_1/k_BT})+\Delta_1^2(1+e^{\Delta_2/k_BT})]
\end{split}
\end{equation}
Using the splitting energy of $\Delta_1 = 112$\,K and $\Delta_2 = 122$\,K from our previously determined CEF parameters, the calculated Schottky anomaly shown in Fig.~\ref{fig:HC}(b) as the black line is well aligned with the data at the higher temperature. This agreement again confirms the validity of our CEF energy scheme deduced from the magnetic susceptibility.

The entropy of the system $S_{4f}$ was obtained by integrating $C_{4f}/T$, and it reaches $R\ln 2$ at $20$\,K, $R\ln 4$ at $60$\,K, $R\ln 6$ at $200$\,K, as shown in Fig.~\ref{fig:HC}(c). The value at $T_\textrm{C}$ is $0.81 R \ln 2$, slightly smaller than $R\ln2$ suggesting the presence of a weak Kondo effect.


\subsection{Raman Spectroscopy}

\begin{figure}[]
\center
\includegraphics[width=\linewidth]{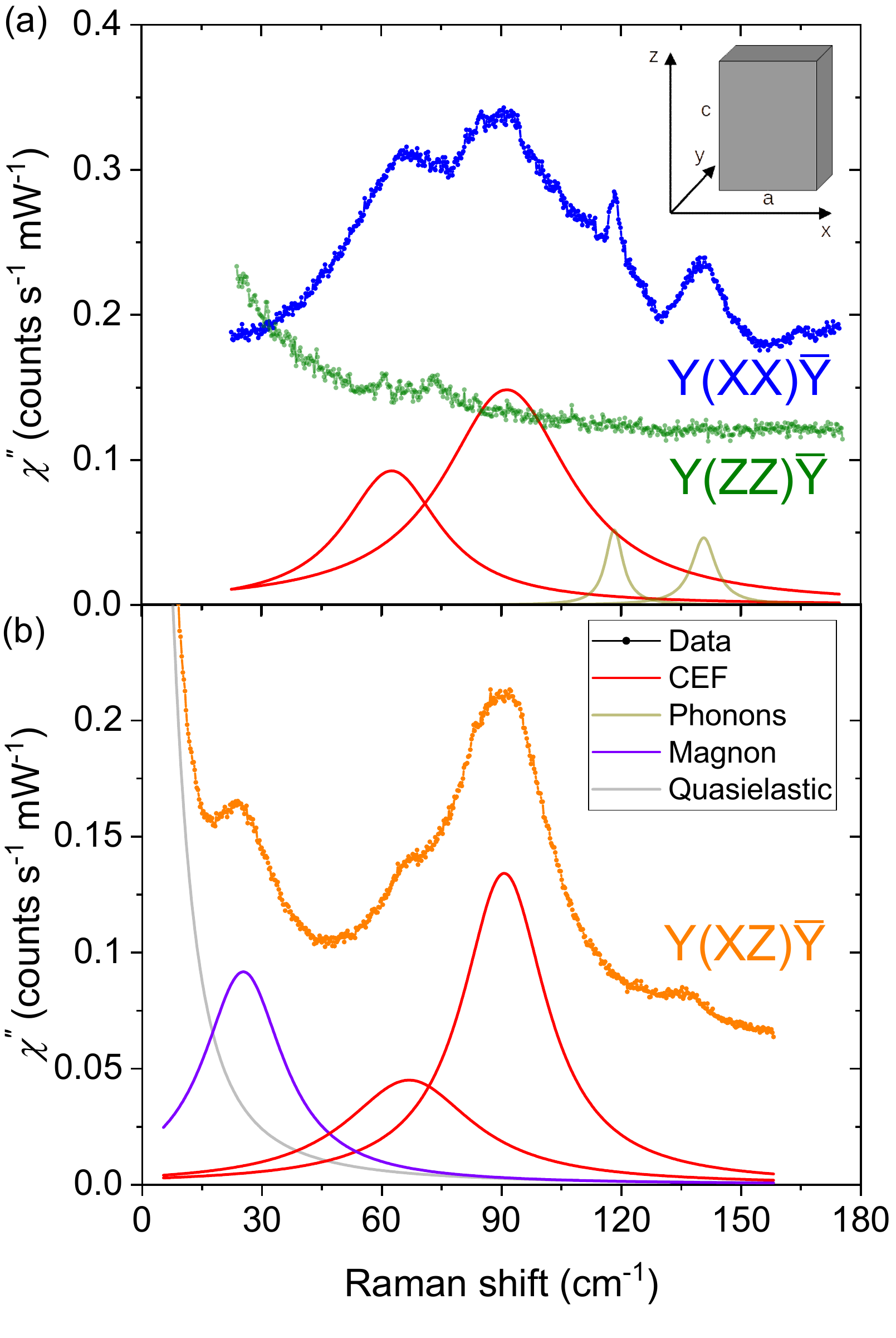}
\caption{(a) Raman spectra of single-crystal CeCuSi with the incident and scattered light parallel to one another. (b) Raman spectrum with perpendicular incident and scattered lights. Solid lines fit to the data. The inset shows the crystal orientation with respect to the optical axes.}
\label{fig:Raman}
\end{figure}

Low-temperature polarized Raman spectroscopy enables the simultaneous measurement of the energies of the CEF excitations and their symmetries. For the CEF scheme depicted in Fig.~\ref{fig:MvsTH}(b) and from group theory, we expect an A$_{1g}\oplus$A$_{2g}\oplus$E$_{g}$ response from the lower-energy transition and an E$_{g}$ response from the higher-energy transition. Figure~\ref{fig:Raman} shows low-energy Raman spectra at 2\,K in various scattering geometries. Table \ref{tab:Symmetries} summarizes the symmetries observable in each geometry, since the global and point groups differ, so do the E-based symmetries for the phonons and CEF modes, respectively. 


\begin{table}[]
\centering
\begin{tabular}{|c|c|c|}
\hline
Porto notation & $D_{6h}$                      & $D_{3d}$             \\ \hline
Y(XX)$\mathrm{\overline{Y}}$         & E$_{2g}\oplus$A$_{1g}$          & E$_{g}\oplus$A$_{1g}$ \\ \hline
Y(ZZ)$\mathrm{\overline{Y}}$         & A$_{1g}$ & A$_{1g}$              \\ \hline
Y(XZ)$\mathrm{\overline{Y}}$         & E$_{1g}$                        & E$_{g}$               \\ \hline
Y(ZX)$\mathrm{\overline{Y}}$         & E$_{1g}$                        & E$_{g}$               \\ \hline
\end{tabular}
\caption{The observable symmetries for a given scattering orientation of light with respect to the global symmetry of the lattice ($D_{6h}$) and the local point symmetry of the Ce$^{3+}$ ions ($D_{3d}$). }
\label{tab:Symmetries}
\end{table}

In Y(XX)$\mathrm{\overline{Y}}$ and Y(XZ)$\mathrm{\overline{Y}}$, we see two strong responses at approximately 65\,{cm$^{-1}$} and 90\,{cm$^{-1}$}, which correspond to 94\,K and 129\,K. Not only do these energies agree reasonably well with the values extracted from the magnetic susceptibility for the CEF levels, but the symmetry dependence, shown in Table \ref{tab:Symmetries}, shows that these responses are of E symmetry. Furthermore, the responses respect the selection rules of the $D_{3d}$ point group, not those of the global symmetry, and hence are the expected E$_g$ responses coming from the CEF excitations. For the first CEF excitation, we also expected to see an A$_{1g}$ response in the Y(ZZ)$\mathrm{\overline{Y}}$ channel, but we did not observe any indication of it. Though group theory predicts that the response exists, it is not always the case that the response is visible, as the Raman matrix element can be intrinsically weak. Generally, we then observe the CEF excitations with the expected symmetry and in an energy range consistent with specific heat and magnetic susceptibility measurements. However, as shown in the Appendix~\ref{AdditionalRaman}, we also measure an additional mode which may be due to magnetoelastic coupling. We also discuss the phonon modes observed by Raman spectroscopy and in relation with the quality of the samples. 

In the Y(XZ)$\mathrm{\overline{Y}}$ channel, shown in Fig.~\ref{fig:Raman}(b), we observe an excitation at 25\,{cm$^{-1}$} (36\,K), which is comparable in energy to the FM magnon gap observed in specific heat. In accordance with Fleury-Loudon theory \cite{FLEURY1968}, the magnon is observed in perpendicular polarization and is consistent with a single-magnon process originating from the acoustic branches at the $\Gamma$ point \cite{Inoue1970}. It is clearly not expected for two-magnon scattering \cite{Inoue1970, Benfatto2006}.

\subsection{Electrical Resistivity}

\begin{figure}[!htb]
\center
\includegraphics[width=\linewidth]{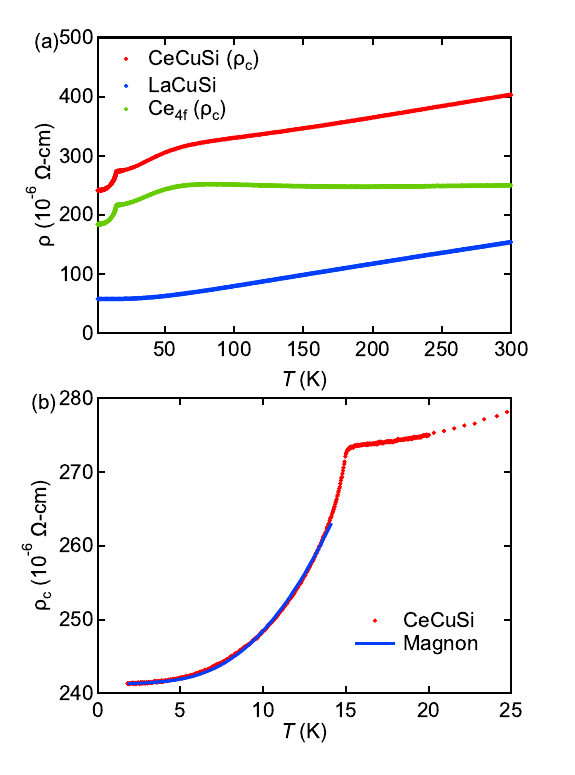}
\caption{(a) The temperature dependence of electrical resistivity $\rho(T)$ of a representative sample of CeCuSi along the $c$ axis, and of a polycrystal of LaCuSi. The $\rho_{4f}$ is also obtained. (b) The low-temperature $\rho(T)$ of CeCuSi. The blue line shows the fitting to the ferromagnetic magnon contribution [Eq.~(\ref{resistFM})].}
\label{fig:resistivity}
\end{figure}

Figure~\ref{fig:resistivity}(a) shows the temperature dependence of electrical resistivity along the $c$ axis of CeCuSi (red curve) and of LaCuSi polycrystalline sample (blue curve). The $4f$ contribution to the resistivity is deduced as shown in the green curve, and it is almost temperature-independent at the higher temperature.

The resistivity data below $T_\textrm{C}$ can be fitted to the FM spin wave gap expression~\cite{andersen1979Electronmagnon}:
\begin{equation} \label{resistFM}
\rho (T) = \rho_{0} + AT^2 + BT\Delta(1 + 2T\Delta^{-1})e^{-\Delta/T},
\end{equation}
where $\rho_{o}$ is the residual resistivity, $A$ is the coefficient for electron-electron scattering, $B$ is the coefficient corresponding to electron-magnon scattering, and $\Delta$ is the energy gap in the magnon excitation~\cite{andersen1979Electronmagnon}. The low-temperature resistivity was first fit to the quadratic Fermi-liquid expression of $\rho (T) = \rho_{0} + AT^2$ from $1.8$\,K to $3$\,K, with $A = 0.01875$\,$\mu\Omega\,\mathrm{cm}\,\mathrm{K}^{-2}$. The ratio of $A$ to the square of the electronic specific-heat coefficient $\gamma^2$ yields $A/\gamma^2 = 11.0$\,$\mu\Omega\,\mathrm{cm}\,\mathrm{mol}^{2}\,\mathrm{K}^{2}\,\mathrm{J}^{-2}$, using the two values found previously from the low-temperature data. This is very close to the universal Kadowaki–Woods ratio of $10$\,$\mu\Omega\,\mathrm{cm}\,\mathrm{mol}^{2}\,\mathrm{K}^{2}\,\mathrm{J}^{-2}$ in many heavy-fermion compounds~\cite{kadowaki1986Universal,jacko2009Unified}.

By fixing the value of $A$, Eq.~\ref{resistFM} gives a good fit up to 14.2\,K as the blue curve shown in Fig~\ref{fig:resistivity}(b). This yields $\rho_{0}=241$~\,$\mu\Omega\,\mathrm{cm}$, $B=0.133$~\,$\mu\Omega\,\mathrm{cm}\,\mathrm{K}^{-2}$, and a magnon gap of $\Delta = 23.9$\,K that is comparable to the value of $25.3$\,K derived from the heat capacity. Surprisingly, the large residual resistivity and low RRR of 1.7 indicates that our single crystal has a lower quality of electronic transport than the polycrystalline sample ($\rho_{0}=9$~\,$\mu\Omega\,\mathrm{cm}$ and RRR$=6.4$~\cite{gignoux1986Magnetic}). Similarly, our polycrystalline LaCuSi sample has a $\rho_{0}=57$~\,$\mu\Omega\,\mathrm{cm}$, which also significantly higher than the $\rho_{0}$ of the previous reported polycrystalline CeCuSi. Our result on single crystals of CeCuSi is consistent over five samples from two different batches. The higher residual resistivity and lower RRR value might be due to the micro-cracks inside the single crystals\modif{. These micro-cracks could result from defects formed during solution growth or from thermal stress during cooling after synthesis,} as mentioned previously.



\section{Conclusion}

In summary, we have successfully synthesized single crystals of CeCuSi by the flux method for the first time. We confirmed that the compound has an ordered $P6_{3}/mmc$ structure through single-crystal and powder XRD. The magnetic susceptibility anisotropy, the Schottky anomaly in heat capacity, and Raman spectra can be well described by the CEF model, and the results agree well with the previous inelastic neutron scattering results. The ferromagnetic magnon behavior has been observed in heat capacity, electrical resistivity, and Raman measurements, further indicating the compound is highly anisotropic.

The results indicate a CEF ground state doublet with magnetic moment primarily in the basal plane with a ferromagnetic interaction. CeCuSi does not exhibit hard-axis ordering, unlike many other FM KL compounds. The interactions along both axes are ferromagnetic, potentially explaining the absence of hard-axis ordering. This underscores the importance of detailed CEF analysis on single crystals to disentangle the CEF contribution from the exchange interaction contribution in the magnetic susceptibility data, thereby advancing our understanding of these materials.

\section{Acknowledgement}

We thank Sophie Tenc\'e for useful discussions. This work is financially supported by the Physics Department, University of California, Davis, U.S.A. We also acknowledge the support from the UC Davis Physics Liquid Helium Laboratory fund. V.T. acknowledges funding from the Laboratoire d'excellence LANEF in Grenoble (ANR-10-LABX-51-01.)   M.-A.M. thanks the European Research Council (ERC) under the European Union's Horizon 2020 research and innovation programme (Grant Agreement No. 865826). We also thank the National Science Foundation (Grant CHE-1531193) for the dual source x-ray diffractometer.

\appendix
\section{Trigonal CEF} \label{Tri_CEF}
For a Ce atom in a trigonal site symmetry, the CEF Hamiltonian can be written as, 
\begin{equation} \label{eq:H}
H_{\mathrm{CEF}}=B_{2}^{0}O_{2}^{0}+B_{4}^{0}O_{4}^{0}+B_{4}^{3}O_{4}^{3},
\end{equation}
where $B_{m}^{n}$ and $O_{m}^{n}$ are the CEF parameters and the Stevens operators, respectively~\cite{stevens1952Matrix,hutchings1964Pointcharge}. The theoretical expression of the magnetic susceptibility with the Van Vleck contribution is given by
\begin{equation} \label{eq:chi}
\begin{aligned}
\chi =\frac{N_{A}g_{J}^{2}\mu _\mathrm{{B}}^{2}\mu_{0} }{\mathcal{Z}}&\Big[\sum_{n}^{}\beta|\braket{J_{i,n}}|^2e^{-\beta \Delta_n}\\ + 2&\sum_{m\neq n}|\Braket{m|J_{i,n}|n}|^2\left(\frac{e^{-\beta \Delta_{m}}-e^{-\beta \Delta_{n}}}{\Delta_{n}-\Delta_{m}}\right)\Big]
\end{aligned}
\end{equation}
where $\beta=1/k_BT$, $\mathcal{Z}=\sum_{n}e^{-\beta \Delta_{n}}$, $i=x,z$, and $n,m=0,1,2$ for the three doublets of Ce$^{3+}$ with $J=5/2$. A detailed derivation of the theoretical magnetic susceptibility expression for the trigonal point symmetry of the Ce atoms can be found in Ref.~\cite{banda2018Crystalline}. Following the same calculation, the final paramagnetic and Van Vleck susceptibilities in both axes in the trigonal point symmetry are given by:
\begin{equation} \label{eq:chipara}
\begin{split}
\chi_{B\parallel z} =\frac{N_{A}g_{J}^{2}\mu _\mathrm{{B}}^{2}\mu_{0}}{\mathcal{Z}} \Bigl\{ \frac{\beta}{4}\Big[(5\cos^2\alpha-\sin^2\alpha)^2+9e^{-\beta \Delta_{1}}\\+(5\sin^2\alpha-\cos^2\alpha)^2e^{-\beta \Delta_{2}}\Big]\\ + \Big[ 18\sin^2\alpha \cos^2\alpha(\frac{1-e^{-\beta \Delta_{2}}}{\Delta_{2}}) \Big] \Bigr\}
\end{split}
\end{equation}

\begin{equation}
\label{eq:chiperp}
\begin{split}
\chi_{B\perp  z} =\frac{N_{A}g_{J}^{2}\mu _\mathrm{{B}}^{2}\mu_{0}}{\mathcal{Z}} \Bigl\{ \frac{9\beta}{4}\Big[\sin^4\alpha+\cos^4\alpha e^{-\beta \Delta_{2}}\Big]\\ + \frac{1}{2}\Big[(5\cos^2\alpha+8\sin^2\alpha)
\frac{1-e^{-\beta \Delta_{1}}}{\Delta_{1}}\\ + (9\sin^2\alpha \cos^2\alpha)\frac{1-e^{-\beta \Delta_{2}}}{\Delta_{2}}\\ + (5\sin^2\alpha+8\cos^2\alpha)\frac{e^{-\beta \Delta_{1}}-e^{-\beta \Delta_{2}}}{\Delta_{2}-\Delta_{1}} \Big] \Bigr\}
\end{split}
\end{equation}
where $\Delta_{1}$ and $\Delta_{2}$ represent the splitting energies. The energy levels and the mixing angle $\alpha$ can be expressed by CEF parameters, and the detailed expression for the trigonal system can be found in Ref.~\cite{banda2018Crystalline}. 

Before doing the fitting to obtain the three CEF parameters, we can estimate the first CEF parameter $B_{2}^{0}$ from the paramagnetic Curie-Weiss temperatures $\theta_{CW}^{\parallel}$ and $\theta_{CW}^{\perp}$ by the following expression~\cite{bowden1971Crystal,wang1971Crystalfield}, which gives the strength of the magnetocrystalline anisotropy:
\begin{equation} \label{B20CW}
B_{2}^{0} = (\theta _{CW}^{\perp}-\theta _{CW}^{\parallel})\frac{10k_{B}}{3(2J-1)(2J+3)}
\end{equation}
Thus the estimated $B_{2}^{0}$ is 4.35\,K, and using this value as the starting point, the inverse magnetic susceptibility including the molecular field contribution $\lambda_{i}$ to account the exchange interactions and the residual susceptibility $\chi_{0}^{i}$ is calculated as~\cite{kabeya2022Eigenstate}:
\begin{equation} \label{eq:inverchi}
\chi_{i}^{-1}=\left ( \frac{\chi_{i}^\mathrm{CEF}(T)}{1-\lambda_{i} \chi_{i}^\mathrm{CEF}(T)} + \chi_{0}^{i} \right )^{-1}
\end{equation}

Under the trigonal symmetry, the states will split into a pure $\ket{\pm 3/2}$ state, and two mixtures of $\ket{\pm 1/2}$ and $\ket{\pm 5/2}$ states.

\section{Additional Raman Spectroscopy Results} \label{AdditionalRaman}

\begin{figure}[]
\center
\includegraphics[width=\linewidth]{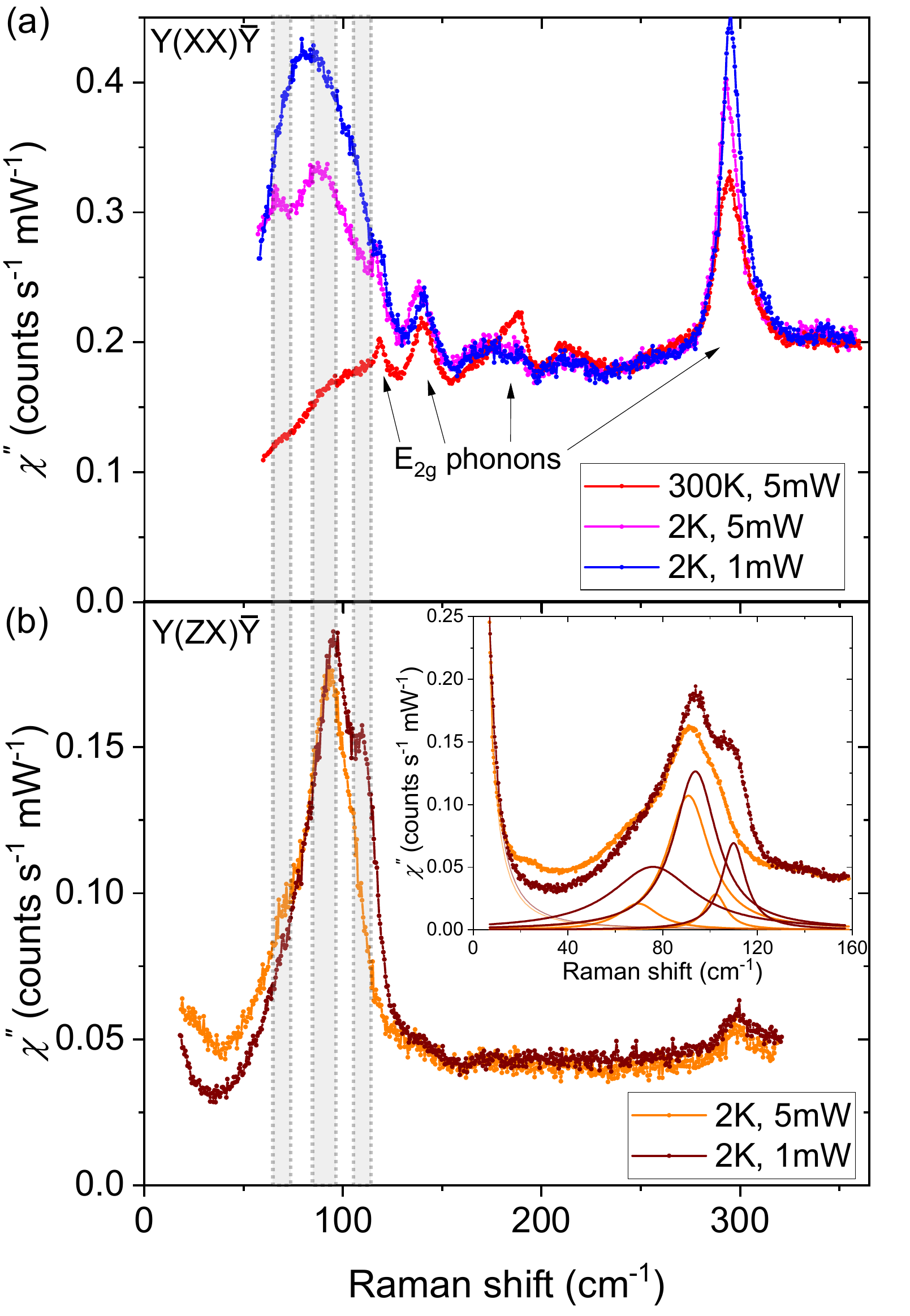}
\caption{(a) Wide energy range Raman spectra of single-crystal CeCuSi with the incident and scattered light parallel to one another at high and low cryostat temperatures and different laser powers. Arrows mark the location of phonon modes. (b) Wide energy range Raman spectra with perpendicular incident and scattered light at low temperatures with different laser powers. Inset shows the higher resolution and lower energy spectra at low temperature with different laser powers. Solid lines show fits to the data at the two laser powers. Translucent boxes show the approximate energies of the CEF transitions across (a) and (b).}
\label{fig:SIRaman}
\end{figure}

From the crystal structure and atomic coordinates, we can predict the number of phonon modes and their symmetries: for CeCuSi, we expect 2E$_{2g}$ modes which should be visible in the Y(XX)$\mathrm{\overline{Y}}$ geometry. At high temperatures, the phonon modes should be the only observable excitations since the CEF levels would be thermally occupied. Figure~\ref{fig:SIRaman}(a) shows a high-temperature Raman spectrum with 4E$_{2g}$ modes indicated, which is unexpected. Although we have synthesized a single crystal of CeCuSi and the SCXRD refinement shows no detectable evidence of mixed occupancy between Cu and Si, it remains conceivable that the similarly sized copper and silicon could have partially interchanged during the growth of the larger crystals used for Raman spectroscopy measurements, producing twice as many phonons with different energies.


Regarding the CEF levels, by reducing the power used and thus the effective temperature of the sample, and in the Y(XX)$\mathrm{\overline{Y}}$ geometry shown in Fig.~\ref{fig:SIRaman}(a), we observe a large increase in the intensity of the CEF response. More notably, in the Y(ZX)$\mathrm{\overline{Y}}$ configuration, Fig.~\ref{fig:SIRaman}(b), we observe that a clear third peak appears at low power. In the inset, higher resolution measurements alongside fittings of the two spectra show that the 5\,mW data are also consistent with three excitations. Within the ferromagnetic phase, a splitting of the ground state could lead to the appearance of the third excitation. However, the internal field is estimated to be 0.225\,T with a 0.25\,{cm$^{-1}$} (0.36\,K) splitting in CEF simulation, which is not sufficient to cause the splitting observed. Due to the close proximity of an E$_{2g}$ phonon, there is potentially magnetoelastic coupling between the phonon and the CEF levels, which could give rise to a hybrid vibronic state, as observed previously in some cerium compounds \cite{Thalmeier1982,Thalmeier1984}.



\begin{thebibliography}{50}%
\makeatletter
\providecommand \@ifxundefined [1]{%
 \@ifx{#1\undefined}
}%
\providecommand \@ifnum [1]{%
 \ifnum #1\expandafter \@firstoftwo
 \else \expandafter \@secondoftwo
 \fi
}%
\providecommand \@ifx [1]{%
 \ifx #1\expandafter \@firstoftwo
 \else \expandafter \@secondoftwo
 \fi
}%
\providecommand \natexlab [1]{#1}%
\providecommand \enquote  [1]{``#1''}%
\providecommand \bibnamefont  [1]{#1}%
\providecommand \bibfnamefont [1]{#1}%
\providecommand \citenamefont [1]{#1}%
\providecommand \href@noop [0]{\@secondoftwo}%
\providecommand \href [0]{\begingroup \@sanitize@url \@href}%
\providecommand \@href[1]{\@@startlink{#1}\@@href}%
\providecommand \@@href[1]{\endgroup#1\@@endlink}%
\providecommand \@sanitize@url [0]{\catcode `\\12\catcode `\$12\catcode
  `\&12\catcode `\#12\catcode `\^12\catcode `\_12\catcode `\%12\relax}%
\providecommand \@@startlink[1]{}%
\providecommand \@@endlink[0]{}%
\providecommand \url  [0]{\begingroup\@sanitize@url \@url }%
\providecommand \@url [1]{\endgroup\@href {#1}{\urlprefix }}%
\providecommand \urlprefix  [0]{URL }%
\providecommand \Eprint [0]{\href }%
\providecommand \doibase [0]{https://doi.org/}%
\providecommand \selectlanguage [0]{\@gobble}%
\providecommand \bibinfo  [0]{\@secondoftwo}%
\providecommand \bibfield  [0]{\@secondoftwo}%
\providecommand \translation [1]{[#1]}%
\providecommand \BibitemOpen [0]{}%
\providecommand \bibitemStop [0]{}%
\providecommand \bibitemNoStop [0]{.\EOS\space}%
\providecommand \EOS [0]{\spacefactor3000\relax}%
\providecommand \BibitemShut  [1]{\csname bibitem#1\endcsname}%
\let\auto@bib@innerbib\@empty
\bibitem [{\citenamefont {Ye}\ \emph {et~al.}(2017)\citenamefont {Ye},
  \citenamefont {Lu}, \citenamefont {Li}, \citenamefont {Nakao}, \citenamefont
  {Yang}, \citenamefont {Tada}, \citenamefont {Kitano},\ and\ \citenamefont
  {Hosono}}]{ye2017Copperbased}%
  \BibitemOpen
  \bibfield  {author} {\bibinfo {author} {\bibfnamefont {T.-N.}\ \bibnamefont
  {Ye}}, \bibinfo {author} {\bibfnamefont {Y.}~\bibnamefont {Lu}}, \bibinfo
  {author} {\bibfnamefont {J.}~\bibnamefont {Li}}, \bibinfo {author}
  {\bibfnamefont {T.}~\bibnamefont {Nakao}}, \bibinfo {author} {\bibfnamefont
  {H.}~\bibnamefont {Yang}}, \bibinfo {author} {\bibfnamefont {T.}~\bibnamefont
  {Tada}}, \bibinfo {author} {\bibfnamefont {M.}~\bibnamefont {Kitano}},\ and\
  \bibinfo {author} {\bibfnamefont {H.}~\bibnamefont {Hosono}},\ }\bibfield
  {title} {\bibinfo {title} {Copper-based intermetallic electride catalyst for
  chemoselective hydrogenation reactions},\ }\href {https://doi.org/doi:
  10.1021/jacs.7b08252} {\bibfield  {journal} {\bibinfo  {journal} {Journal of
  the American Chemical Society}\ }\textbf {\bibinfo {volume} {139}},\ \bibinfo
  {pages} {17089} (\bibinfo {year} {2017})}\BibitemShut {NoStop}%
\bibitem [{\citenamefont {Gupta}\ and\ \citenamefont
  {Suresh}(2015)}]{gupta2015Review}%
  \BibitemOpen
  \bibfield  {author} {\bibinfo {author} {\bibfnamefont {S.}~\bibnamefont
  {Gupta}}\ and\ \bibinfo {author} {\bibfnamefont {K.}~\bibnamefont {Suresh}},\
  }\bibfield  {title} {\bibinfo {title} {Review on magnetic and related
  properties of {{R}}{{{\emph{TX}}}} compounds},\ }\href
  {https://doi.org/10.1016/j.jallcom.2014.08.079} {\bibfield  {journal}
  {\bibinfo  {journal} {Journal of Alloys and Compounds}\ }\textbf {\bibinfo
  {volume} {618}},\ \bibinfo {pages} {562} (\bibinfo {year}
  {2015})}\BibitemShut {NoStop}%
\bibitem [{\citenamefont {P{\"o}ttgen}\ and\ \citenamefont
  {Chevalier}(2015{\natexlab{a}})}]{pottgen2015Cerium}%
  \BibitemOpen
  \bibfield  {author} {\bibinfo {author} {\bibfnamefont {R.}~\bibnamefont
  {P{\"o}ttgen}}\ and\ \bibinfo {author} {\bibfnamefont {B.}~\bibnamefont
  {Chevalier}},\ }\bibfield  {title} {\bibinfo {title} {Cerium intermetallics
  with {{ZrNiAl-type}} structure -- a review},\ }\href
  {https://doi.org/doi:10.1515/znb-2015-0018} {\bibfield  {journal} {\bibinfo
  {journal} {Zeitschrift f{\"u}r Naturforschung B}\ }\textbf {\bibinfo {volume}
  {70}},\ \bibinfo {pages} {289} (\bibinfo {year}
  {2015}{\natexlab{a}})}\BibitemShut {NoStop}%
\bibitem [{\citenamefont {P{\"o}ttgen}\ and\ \citenamefont
  {Chevalier}(2015{\natexlab{b}})}]{pottgen2015Equiatomic}%
  \BibitemOpen
  \bibfield  {author} {\bibinfo {author} {\bibfnamefont {R.}~\bibnamefont
  {P{\"o}ttgen}}\ and\ \bibinfo {author} {\bibfnamefont {B.}~\bibnamefont
  {Chevalier}},\ }\bibfield  {title} {\bibinfo {title} {Equiatomic cerium
  intermetallics {{CeXX}}{\textsuperscript{{$\prime$}}} with two p elements},\
  }\href {https://doi.org/doi:10.1515/znb-2015-0109} {\bibfield  {journal}
  {\bibinfo  {journal} {Zeitschrift f{\"u}r Naturforschung B}\ }\textbf
  {\bibinfo {volume} {70}},\ \bibinfo {pages} {695} (\bibinfo {year}
  {2015}{\natexlab{b}})}\BibitemShut {NoStop}%
\bibitem [{\citenamefont {P{\"o}ttgen}\ \emph {et~al.}(2016)\citenamefont
  {P{\"o}ttgen}, \citenamefont {Janka},\ and\ \citenamefont
  {Chevalier}}]{pottgen2016Cerium}%
  \BibitemOpen
  \bibfield  {author} {\bibinfo {author} {\bibfnamefont {R.}~\bibnamefont
  {P{\"o}ttgen}}, \bibinfo {author} {\bibfnamefont {O.}~\bibnamefont {Janka}},\
  and\ \bibinfo {author} {\bibfnamefont {B.}~\bibnamefont {Chevalier}},\
  }\bibfield  {title} {\bibinfo {title} {Cerium intermetallics
  {{Ce}}{{{\emph{TX}}}} -- review {{III}}},\ }\href
  {https://doi.org/doi:10.1515/znb-2016-0013} {\bibfield  {journal} {\bibinfo
  {journal} {Zeitschrift f{\"u}r Naturforschung B}\ }\textbf {\bibinfo {volume}
  {71}},\ \bibinfo {pages} {165} (\bibinfo {year} {2016})}\BibitemShut
  {NoStop}%
\bibitem [{\citenamefont {Janka}\ \emph {et~al.}(2016)\citenamefont {Janka},
  \citenamefont {Niehaus}, \citenamefont {P{\"o}ttgen},\ and\ \citenamefont
  {Chevalier}}]{janka2016Cerium}%
  \BibitemOpen
  \bibfield  {author} {\bibinfo {author} {\bibfnamefont {O.}~\bibnamefont
  {Janka}}, \bibinfo {author} {\bibfnamefont {O.}~\bibnamefont {Niehaus}},
  \bibinfo {author} {\bibfnamefont {R.}~\bibnamefont {P{\"o}ttgen}},\ and\
  \bibinfo {author} {\bibfnamefont {B.}~\bibnamefont {Chevalier}},\ }\bibfield
  {title} {\bibinfo {title} {Cerium intermetallics with {{TiNiSi-type}}
  structure},\ }\href {https://doi.org/doi:10.1515/znb-2016-0101} {\bibfield
  {journal} {\bibinfo  {journal} {Zeitschrift f{\"u}r Naturforschung B}\
  }\textbf {\bibinfo {volume} {71}},\ \bibinfo {pages} {737} (\bibinfo {year}
  {2016})}\BibitemShut {NoStop}%
\bibitem [{\citenamefont {Balents}(2010)}]{balents2010Spin}%
  \BibitemOpen
  \bibfield  {author} {\bibinfo {author} {\bibfnamefont {L.}~\bibnamefont
  {Balents}},\ }\bibfield  {title} {\bibinfo {title} {Spin liquids in
  frustrated magnets},\ }\href {https://doi.org/10.1038/nature08917} {\bibfield
   {journal} {\bibinfo  {journal} {Nature}\ }\textbf {\bibinfo {volume}
  {464}},\ \bibinfo {pages} {199} (\bibinfo {year} {2010})}\BibitemShut
  {NoStop}%
\bibitem [{\citenamefont {Huang}\ \emph {et~al.}(2014)\citenamefont {Huang},
  \citenamefont {Chen},\ and\ \citenamefont {Hermele}}]{huang2014Quantum}%
  \BibitemOpen
  \bibfield  {author} {\bibinfo {author} {\bibfnamefont {Y.-P.}\ \bibnamefont
  {Huang}}, \bibinfo {author} {\bibfnamefont {G.}~\bibnamefont {Chen}},\ and\
  \bibinfo {author} {\bibfnamefont {M.}~\bibnamefont {Hermele}},\ }\bibfield
  {title} {\bibinfo {title} {Quantum spin ices and topological phases from
  dipolar-octupolar doublets on the pyrochlore lattice},\ }\href
  {https://doi.org/10.1103/PhysRevLett.112.167203} {\bibfield  {journal}
  {\bibinfo  {journal} {Physical Review Letters}\ }\textbf {\bibinfo {volume}
  {112}},\ \bibinfo {pages} {167203} (\bibinfo {year} {2014})}\BibitemShut
  {NoStop}%
\bibitem [{\citenamefont {Sibille}\ \emph {et~al.}(2015)\citenamefont
  {Sibille}, \citenamefont {Lhotel}, \citenamefont {Pomjakushin}, \citenamefont
  {Baines}, \citenamefont {Fennell},\ and\ \citenamefont
  {Kenzelmann}}]{sibille2015Candidate}%
  \BibitemOpen
  \bibfield  {author} {\bibinfo {author} {\bibfnamefont {R.}~\bibnamefont
  {Sibille}}, \bibinfo {author} {\bibfnamefont {E.}~\bibnamefont {Lhotel}},
  \bibinfo {author} {\bibfnamefont {V.}~\bibnamefont {Pomjakushin}}, \bibinfo
  {author} {\bibfnamefont {C.}~\bibnamefont {Baines}}, \bibinfo {author}
  {\bibfnamefont {T.}~\bibnamefont {Fennell}},\ and\ \bibinfo {author}
  {\bibfnamefont {M.}~\bibnamefont {Kenzelmann}},\ }\bibfield  {title}
  {\bibinfo {title} {Candidate {{Quantum Spin Liquid}} in the
  {{Ce}}{\textsuperscript{3+}} {{Pyrochlore Stannate
  Ce}}{\textsubscript{2}}{{Sn}}{\textsubscript{2}}{{O}}{\textsubscript{7}}},\
  }\href {https://doi.org/10.1103/PhysRevLett.115.097202} {\bibfield  {journal}
  {\bibinfo  {journal} {Physical Review Letters}\ }\textbf {\bibinfo {volume}
  {115}},\ \bibinfo {pages} {097202} (\bibinfo {year} {2015})}\BibitemShut
  {NoStop}%
\bibitem [{\citenamefont {Kirkpatrick}(1977)}]{kirkpatrick1977Frustration}%
  \BibitemOpen
  \bibfield  {author} {\bibinfo {author} {\bibfnamefont {S.}~\bibnamefont
  {Kirkpatrick}},\ }\bibfield  {title} {\bibinfo {title} {Frustration and
  ground-state degeneracy in spin glasses},\ }\href
  {https://doi.org/10.1103/PhysRevB.16.4630} {\bibfield  {journal} {\bibinfo
  {journal} {Physical Review B}\ }\textbf {\bibinfo {volume} {16}},\ \bibinfo
  {pages} {4630} (\bibinfo {year} {1977})}\BibitemShut {NoStop}%
\bibitem [{\citenamefont {Uzoh}\ \emph {et~al.}(2023)\citenamefont {Uzoh},
  \citenamefont {Kim},\ and\ \citenamefont {Mun}}]{uzoh2023Influence}%
  \BibitemOpen
  \bibfield  {author} {\bibinfo {author} {\bibfnamefont {O.~P.}\ \bibnamefont
  {Uzoh}}, \bibinfo {author} {\bibfnamefont {S.}~\bibnamefont {Kim}},\ and\
  \bibinfo {author} {\bibfnamefont {E.}~\bibnamefont {Mun}},\ }\bibfield
  {title} {\bibinfo {title} {Influence of crystalline electric field on the
  magnetic properties of {{CeCd}}{\textsubscript{3}}{{X}}{\textsubscript{3}}
  ({{X}} = {{P}}, {{As}})},\ }\href
  {https://doi.org/10.1103/PhysRevMaterials.7.013402} {\bibfield  {journal}
  {\bibinfo  {journal} {Physical Review Materials}\ }\textbf {\bibinfo {volume}
  {7}},\ \bibinfo {pages} {013402} (\bibinfo {year} {2023})}\BibitemShut
  {NoStop}%
\bibitem [{\citenamefont {Ochiai}\ \emph {et~al.}(2021)\citenamefont {Ochiai},
  \citenamefont {Kabeya}, \citenamefont {Maniwa}, \citenamefont {Saito},
  \citenamefont {Nakamura},\ and\ \citenamefont
  {Katoh}}]{ochiai2021Fieldinduced}%
  \BibitemOpen
  \bibfield  {author} {\bibinfo {author} {\bibfnamefont {A.}~\bibnamefont
  {Ochiai}}, \bibinfo {author} {\bibfnamefont {N.}~\bibnamefont {Kabeya}},
  \bibinfo {author} {\bibfnamefont {K.}~\bibnamefont {Maniwa}}, \bibinfo
  {author} {\bibfnamefont {M.}~\bibnamefont {Saito}}, \bibinfo {author}
  {\bibfnamefont {S.}~\bibnamefont {Nakamura}},\ and\ \bibinfo {author}
  {\bibfnamefont {K.}~\bibnamefont {Katoh}},\ }\bibfield  {title} {\bibinfo
  {title} {Field-induced anomalous magnetic state beyond the magnetically
  ordered state in the slightly distorted triangular {{S}}=1/2 rare-earth
  antiferromagnet {{CeZn}}{\textsubscript{3}}{{P}}{\textsubscript{3}}},\ }\href
  {https://doi.org/10.1103/PhysRevB.104.144420} {\bibfield  {journal} {\bibinfo
   {journal} {Physical Review B}\ }\textbf {\bibinfo {volume} {104}},\ \bibinfo
  {pages} {144420} (\bibinfo {year} {2021})}\BibitemShut {NoStop}%
\bibitem [{\citenamefont {Rieger}\ and\ \citenamefont
  {Parth{\'e}}(1969)}]{rieger1969Ternare}%
  \BibitemOpen
  \bibfield  {author} {\bibinfo {author} {\bibfnamefont {W.}~\bibnamefont
  {Rieger}}\ and\ \bibinfo {author} {\bibfnamefont {E.}~\bibnamefont
  {Parth{\'e}}},\ }\bibfield  {title} {\bibinfo {title} {Tern{\"a}re
  erdalkali-und seltene erd-silicide und-{{Germanide}} mit
  {{AlB}}{\textsubscript{2}}-{{Struktur}}},\ }\href
  {https://doi.org/10.1007/BF00904085} {\bibfield  {journal} {\bibinfo
  {journal} {Monatshefte f{\"u}r Chemie / Chemical Monthly}\ }\textbf {\bibinfo
  {volume} {100}},\ \bibinfo {pages} {439} (\bibinfo {year}
  {1969})}\BibitemShut {NoStop}%
\bibitem [{\citenamefont {Gignoux}\ \emph {et~al.}(1986)\citenamefont
  {Gignoux}, \citenamefont {Schmitt},\ and\ \citenamefont
  {Zerguine}}]{gignoux1986Magnetic}%
  \BibitemOpen
  \bibfield  {author} {\bibinfo {author} {\bibfnamefont {D.}~\bibnamefont
  {Gignoux}}, \bibinfo {author} {\bibfnamefont {D.}~\bibnamefont {Schmitt}},\
  and\ \bibinfo {author} {\bibfnamefont {M.}~\bibnamefont {Zerguine}},\
  }\bibfield  {title} {\bibinfo {title} {Magnetic properties of {{CeCuSi}}},\
  }\href {https://doi.org/10.1016/0038-1098(86)90796-9} {\bibfield  {journal}
  {\bibinfo  {journal} {Solid State Communications}\ }\textbf {\bibinfo
  {volume} {58}},\ \bibinfo {pages} {559} (\bibinfo {year} {1986})}\BibitemShut
  {NoStop}%
\bibitem [{\citenamefont {Iandelli}(1983)}]{iandelli1983Low}%
  \BibitemOpen
  \bibfield  {author} {\bibinfo {author} {\bibfnamefont {A.}~\bibnamefont
  {Iandelli}},\ }\bibfield  {title} {\bibinfo {title} {A low temperature
  crystal modification of the rare earth ternary compounds {{RCuSi}}},\ }\href
  {https://doi.org/10.1016/0022-5088(83)90123-6} {\bibfield  {journal}
  {\bibinfo  {journal} {Journal of the Less Common Metals}\ }\textbf {\bibinfo
  {volume} {90}},\ \bibinfo {pages} {121} (\bibinfo {year} {1983})}\BibitemShut
  {NoStop}%
\bibitem [{\citenamefont {{Sondezi-Mhlungu}}\ \emph {et~al.}(2009)\citenamefont
  {{Sondezi-Mhlungu}}, \citenamefont {Adroja}, \citenamefont {Strydom},
  \citenamefont {Paschen},\ and\ \citenamefont
  {Goremychkin}}]{sondezi-mhlungu2009Crystal}%
  \BibitemOpen
  \bibfield  {author} {\bibinfo {author} {\bibfnamefont {B.}~\bibnamefont
  {{Sondezi-Mhlungu}}}, \bibinfo {author} {\bibfnamefont {D.}~\bibnamefont
  {Adroja}}, \bibinfo {author} {\bibfnamefont {A.}~\bibnamefont {Strydom}},
  \bibinfo {author} {\bibfnamefont {S.}~\bibnamefont {Paschen}},\ and\ \bibinfo
  {author} {\bibfnamefont {E.}~\bibnamefont {Goremychkin}},\ }\bibfield
  {title} {\bibinfo {title} {Crystal electric field excitations in
  ferromagnetic {{Ce}}{{{\emph{TX}}}} compounds},\ }\href
  {https://doi.org/10.1016/j.physb.2009.07.014} {\bibfield  {journal} {\bibinfo
   {journal} {Physica B: Condensed Matter}\ }\textbf {\bibinfo {volume}
  {404}},\ \bibinfo {pages} {3032} (\bibinfo {year} {2009})}\BibitemShut
  {NoStop}%
\bibitem [{\citenamefont {Yang}\ \emph {et~al.}(1991)\citenamefont {Yang},
  \citenamefont {Kuang}, \citenamefont {Li}, \citenamefont {Br{\"u}ck},
  \citenamefont {Nakotte}, \citenamefont {{de Boer}}, \citenamefont {Wu},
  \citenamefont {Li},\ and\ \citenamefont {Wang}}]{yang1991Magnetic}%
  \BibitemOpen
  \bibfield  {author} {\bibinfo {author} {\bibfnamefont {F.}~\bibnamefont
  {Yang}}, \bibinfo {author} {\bibfnamefont {J.~P.}\ \bibnamefont {Kuang}},
  \bibinfo {author} {\bibfnamefont {J.}~\bibnamefont {Li}}, \bibinfo {author}
  {\bibfnamefont {E.}~\bibnamefont {Br{\"u}ck}}, \bibinfo {author}
  {\bibfnamefont {H.}~\bibnamefont {Nakotte}}, \bibinfo {author} {\bibfnamefont
  {F.~R.}\ \bibnamefont {{de Boer}}}, \bibinfo {author} {\bibfnamefont
  {X.}~\bibnamefont {Wu}}, \bibinfo {author} {\bibfnamefont {Z.}~\bibnamefont
  {Li}},\ and\ \bibinfo {author} {\bibfnamefont {Y.}~\bibnamefont {Wang}},\
  }\bibfield  {title} {\bibinfo {title} {Magnetic properties of {{CeCuX}}
  compounds},\ }\href {https://doi.org/10.1063/1.348279} {\bibfield  {journal}
  {\bibinfo  {journal} {Journal of Applied Physics}\ }\textbf {\bibinfo
  {volume} {69}},\ \bibinfo {pages} {4705} (\bibinfo {year}
  {1991})}\BibitemShut {NoStop}%
\bibitem [{\citenamefont {Hearne}\ \emph {et~al.}(2014)\citenamefont {Hearne},
  \citenamefont {Diguet}, \citenamefont {Strydom}, \citenamefont
  {{Sondezi-Mhlungu}}, \citenamefont {K.~Kamenev},\ and\ \citenamefont
  {Nataf}}]{hearne2014Pressure}%
  \BibitemOpen
  \bibfield  {author} {\bibinfo {author} {\bibfnamefont {G.~R.}\ \bibnamefont
  {Hearne}}, \bibinfo {author} {\bibfnamefont {G.}~\bibnamefont {Diguet}},
  \bibinfo {author} {\bibfnamefont {A.~M.}\ \bibnamefont {Strydom}}, \bibinfo
  {author} {\bibfnamefont {B.}~\bibnamefont {{Sondezi-Mhlungu}}}, \bibinfo
  {author} {\bibfnamefont {F.~B.}\ \bibnamefont {K.~Kamenev}},\ and\ \bibinfo
  {author} {\bibfnamefont {L.}~\bibnamefont {Nataf}},\ }\bibfield  {title}
  {\bibinfo {title} {Pressure effects on the magnetic behavior of the local
  moment ferromagnet {{CeCuSi}}},\ }\href{http://events.saip.org.za/getFile.py/access?resId=7&materialId=6&confId=34} {\bibfield  {journal}
  {\bibinfo  {journal} {in Proceedings of SAIP2014, the 59th Annual Conference
  of the South African Institute of Physics}\ } (\bibinfo {year}
  {2014})}\BibitemShut {NoStop}%
\bibitem [{\citenamefont {Hafner}\ \emph {et~al.}(2019)\citenamefont {Hafner},
  \citenamefont {Rai}, \citenamefont {Banda}, \citenamefont {Kliemt},
  \citenamefont {Krellner}, \citenamefont {Sichelschmidt}, \citenamefont
  {Morosan}, \citenamefont {Geibel},\ and\ \citenamefont
  {Brando}}]{hafner2019Kondolattice}%
  \BibitemOpen
  \bibfield  {author} {\bibinfo {author} {\bibfnamefont {D.}~\bibnamefont
  {Hafner}}, \bibinfo {author} {\bibfnamefont {B.~K.}\ \bibnamefont {Rai}},
  \bibinfo {author} {\bibfnamefont {J.}~\bibnamefont {Banda}}, \bibinfo
  {author} {\bibfnamefont {K.}~\bibnamefont {Kliemt}}, \bibinfo {author}
  {\bibfnamefont {C.}~\bibnamefont {Krellner}}, \bibinfo {author}
  {\bibfnamefont {J.}~\bibnamefont {Sichelschmidt}}, \bibinfo {author}
  {\bibfnamefont {E.}~\bibnamefont {Morosan}}, \bibinfo {author} {\bibfnamefont
  {C.}~\bibnamefont {Geibel}},\ and\ \bibinfo {author} {\bibfnamefont
  {M.}~\bibnamefont {Brando}},\ }\bibfield  {title} {\bibinfo {title}
  {Kondo-lattice ferromagnets and their peculiar order along the magnetically
  hard axis determined by the crystalline electric field},\ }\href
  {https://doi.org/10.1103/PhysRevB.99.201109} {\bibfield  {journal} {\bibinfo
  {journal} {Physical Review B}\ }\textbf {\bibinfo {volume} {99}},\ \bibinfo
  {pages} {201109} (\bibinfo {year} {2019})}\BibitemShut {NoStop}%
\bibitem [{\citenamefont {Jesche}\ and\ \citenamefont
  {Canfield}(2014)}]{jesche2014Single}%
  \BibitemOpen
  \bibfield  {author} {\bibinfo {author} {\bibfnamefont {A.}~\bibnamefont
  {Jesche}}\ and\ \bibinfo {author} {\bibfnamefont {P.}~\bibnamefont
  {Canfield}},\ }\bibfield  {title} {\bibinfo {title} {Single crystal growth
  from light, volatile and reactive materials using lithium and calcium flux},\
  }\href {https://doi.org/10.1080/14786435.2014.913114} {\bibfield  {journal}
  {\bibinfo  {journal} {Philosophical Magazine}\ }\textbf {\bibinfo {volume}
  {94}},\ \bibinfo {pages} {2372} (\bibinfo {year} {2014})}\BibitemShut
  {NoStop}%
\bibitem [{\citenamefont {Toby}\ and\ \citenamefont
  {Von~Dreele}(2013)}]{toby2013GSASII}%
  \BibitemOpen
  \bibfield  {author} {\bibinfo {author} {\bibfnamefont {B.~H.}\ \bibnamefont
  {Toby}}\ and\ \bibinfo {author} {\bibfnamefont {R.~B.}\ \bibnamefont
  {Von~Dreele}},\ }\bibfield  {title} {\bibinfo {title} {{{GSAS-II}}: The
  genesis of a modern open-source all purpose crystallography software
  package},\ }\href {https://doi.org/10.1107/S0021889813003531} {\bibfield
  {journal} {\bibinfo  {journal} {Journal of Applied Crystallography}\ }\textbf
  {\bibinfo {volume} {46}},\ \bibinfo {pages} {544} (\bibinfo {year}
  {2013})}\BibitemShut {NoStop}%
\bibitem [{\citenamefont {Mugnoli}\ \emph {et~al.}(1984)\citenamefont
  {Mugnoli}, \citenamefont {Albinati},\ and\ \citenamefont
  {Hewat}}]{mugnoli1984Neutron}%
  \BibitemOpen
  \bibfield  {author} {\bibinfo {author} {\bibfnamefont {A.}~\bibnamefont
  {Mugnoli}}, \bibinfo {author} {\bibfnamefont {A.}~\bibnamefont {Albinati}},\
  and\ \bibinfo {author} {\bibfnamefont {A.}~\bibnamefont {Hewat}},\ }\bibfield
   {title} {\bibinfo {title} {A neutron powder diffraction study of the crystal
  structure of {{LaCuSi}}},\ }\href
  {https://doi.org/10.1016/0022-5088(84)90041-9} {\bibfield  {journal}
  {\bibinfo  {journal} {Journal of the Less Common Metals}\ }\textbf {\bibinfo
  {volume} {97}},\ \bibinfo {pages} {L1} (\bibinfo {year} {1984})}\BibitemShut
  {NoStop}%
\bibitem [{\citenamefont {Ullah}(2024)}]{ullah2024Experimental}%
  \BibitemOpen
  \bibfield  {author} {\bibinfo {author} {\bibfnamefont {R.}~\bibnamefont
  {Ullah}},\ }\emph {\bibinfo {title} {}}\ \href{https://escholarship.org/uc/item/83b7r66d} {Doctoral dissertation},\ \bibinfo
  {school} {University of California, Davis}, \bibinfo {year}
  {2024}\BibitemShut {NoStop}%
\bibitem [{\citenamefont {Ajeesh}\ \emph {et~al.}(2017)\citenamefont {Ajeesh},
  \citenamefont {Shang}, \citenamefont {Jiang}, \citenamefont {Xie},
  \citenamefont {{dos Reis}}, \citenamefont {Smidman}, \citenamefont {Geibel},
  \citenamefont {Yuan},\ and\ \citenamefont {Nicklas}}]{ajeesh2017Isingtype}%
  \BibitemOpen
  \bibfield  {author} {\bibinfo {author} {\bibfnamefont {M.~O.}\ \bibnamefont
  {Ajeesh}}, \bibinfo {author} {\bibfnamefont {T.}~\bibnamefont {Shang}},
  \bibinfo {author} {\bibfnamefont {W.~B.}\ \bibnamefont {Jiang}}, \bibinfo
  {author} {\bibfnamefont {W.}~\bibnamefont {Xie}}, \bibinfo {author}
  {\bibfnamefont {R.~D.}\ \bibnamefont {{dos Reis}}}, \bibinfo {author}
  {\bibfnamefont {M.}~\bibnamefont {Smidman}}, \bibinfo {author} {\bibfnamefont
  {C.}~\bibnamefont {Geibel}}, \bibinfo {author} {\bibfnamefont {H.~Q.}\
  \bibnamefont {Yuan}},\ and\ \bibinfo {author} {\bibfnamefont
  {M.}~\bibnamefont {Nicklas}},\ }\bibfield  {title} {\bibinfo {title}
  {Ising-type magnetic anisotropy in {{CePd}}{$_{2}$}{{As}}{$_{2}$}},\ }\href
  {https://doi.org/10.1038/s41598-017-07595-w} {\bibfield  {journal} {\bibinfo
  {journal} {Scientific Reports}\ }\textbf {\bibinfo {volume} {7}},\ \bibinfo
  {pages} {7338} (\bibinfo {year} {2017})}\BibitemShut {NoStop}%
\bibitem [{\citenamefont {Huang}\ \emph {et~al.}(2015)\citenamefont {Huang},
  \citenamefont {Fritsch}, \citenamefont {Pilawa}, \citenamefont {Yang},
  \citenamefont {Merz},\ and\ \citenamefont
  {v.~L{\"o}hneysen}}]{huang2015Lowtemperature}%
  \BibitemOpen
  \bibfield  {author} {\bibinfo {author} {\bibfnamefont {C.~L.}\ \bibnamefont
  {Huang}}, \bibinfo {author} {\bibfnamefont {V.}~\bibnamefont {Fritsch}},
  \bibinfo {author} {\bibfnamefont {B.}~\bibnamefont {Pilawa}}, \bibinfo
  {author} {\bibfnamefont {C.~C.}\ \bibnamefont {Yang}}, \bibinfo {author}
  {\bibfnamefont {M.}~\bibnamefont {Merz}},\ and\ \bibinfo {author}
  {\bibfnamefont {H.}~\bibnamefont {v.~L{\"o}hneysen}},\ }\bibfield  {title}
  {\bibinfo {title} {Low-temperature magnetic, thermodynamic, and transport
  properties of antiferromagnetic {{CeAuSn}} single crystals},\ }\href
  {https://doi.org/10.1103/PhysRevB.91.144413} {\bibfield  {journal} {\bibinfo
  {journal} {Physical Review B}\ }\textbf {\bibinfo {volume} {91}},\ \bibinfo
  {pages} {144413} (\bibinfo {year} {2015})}\BibitemShut {NoStop}%
\bibitem [{\citenamefont {Banda}\ \emph {et~al.}(2018)\citenamefont {Banda},
  \citenamefont {Rai}, \citenamefont {Rosner}, \citenamefont {Morosan},
  \citenamefont {Geibel},\ and\ \citenamefont {Brando}}]{banda2018Crystalline}%
  \BibitemOpen
  \bibfield  {author} {\bibinfo {author} {\bibfnamefont {J.}~\bibnamefont
  {Banda}}, \bibinfo {author} {\bibfnamefont {B.~K.}\ \bibnamefont {Rai}},
  \bibinfo {author} {\bibfnamefont {H.}~\bibnamefont {Rosner}}, \bibinfo
  {author} {\bibfnamefont {E.}~\bibnamefont {Morosan}}, \bibinfo {author}
  {\bibfnamefont {C.}~\bibnamefont {Geibel}},\ and\ \bibinfo {author}
  {\bibfnamefont {M.}~\bibnamefont {Brando}},\ }\bibfield  {title} {\bibinfo
  {title} {Crystalline electric field of {{Ce}} in trigonal symmetry:
  {{CeIr}}{$_{3}$}{{Ge}}{$_7$} as a model case},\ }\href
  {https://doi.org/10.1103/PhysRevB.98.195120} {\bibfield  {journal} {\bibinfo
  {journal} {Physical Review B}\ }\textbf {\bibinfo {volume} {98}},\ \bibinfo
  {pages} {195120} (\bibinfo {year} {2018})}\BibitemShut {NoStop}%
\bibitem [{\citenamefont {Rai}\ \emph {et~al.}(2018)\citenamefont {Rai},
  \citenamefont {Banda}, \citenamefont {Stavinoha}, \citenamefont {Borth},
  \citenamefont {Jang}, \citenamefont {Benavides}, \citenamefont {Sokolov},
  \citenamefont {Chan}, \citenamefont {Nicklas}, \citenamefont {Brando},
  \citenamefont {Huang},\ and\ \citenamefont {Morosan}}]{rai2018Mathrm}%
  \BibitemOpen
  \bibfield  {author} {\bibinfo {author} {\bibfnamefont {B.~K.}\ \bibnamefont
  {Rai}}, \bibinfo {author} {\bibfnamefont {J.}~\bibnamefont {Banda}}, \bibinfo
  {author} {\bibfnamefont {M.}~\bibnamefont {Stavinoha}}, \bibinfo {author}
  {\bibfnamefont {R.}~\bibnamefont {Borth}}, \bibinfo {author} {\bibfnamefont
  {D.-J.}\ \bibnamefont {Jang}}, \bibinfo {author} {\bibfnamefont {K.~A.}\
  \bibnamefont {Benavides}}, \bibinfo {author} {\bibfnamefont {D.~A.}\
  \bibnamefont {Sokolov}}, \bibinfo {author} {\bibfnamefont {J.~Y.}\
  \bibnamefont {Chan}}, \bibinfo {author} {\bibfnamefont {M.}~\bibnamefont
  {Nicklas}}, \bibinfo {author} {\bibfnamefont {M.}~\bibnamefont {Brando}},
  \bibinfo {author} {\bibfnamefont {C.-L.}\ \bibnamefont {Huang}},\ and\
  \bibinfo {author} {\bibfnamefont {E.}~\bibnamefont {Morosan}},\ }\bibfield
  {title} {\bibinfo {title} {{{CeIr}}{$_{3}$}{{Ge}}{$_7$}: {{A}} local moment
  antiferromagnetic metal with extremely low ordering temperature},\ }\href
  {https://doi.org/10.1103/PhysRevB.98.195119} {\bibfield  {journal} {\bibinfo
  {journal} {Physical Review B}\ }\textbf {\bibinfo {volume} {98}},\ \bibinfo
  {pages} {195119} (\bibinfo {year} {2018})}\BibitemShut {NoStop}%
\bibitem [{\citenamefont {Dunsiger}\ \emph {et~al.}(2020)\citenamefont
  {Dunsiger}, \citenamefont {Lee}, \citenamefont {Sonier},\ and\ \citenamefont
  {Mun}}]{dunsiger2020Longrange}%
  \BibitemOpen
  \bibfield  {author} {\bibinfo {author} {\bibfnamefont {S.~R.}\ \bibnamefont
  {Dunsiger}}, \bibinfo {author} {\bibfnamefont {J.}~\bibnamefont {Lee}},
  \bibinfo {author} {\bibfnamefont {J.~E.}\ \bibnamefont {Sonier}},\ and\
  \bibinfo {author} {\bibfnamefont {E.~D.}\ \bibnamefont {Mun}},\ }\bibfield
  {title} {\bibinfo {title} {Long-range magnetic order in the anisotropic
  triangular lattice system
  {{CeCd}}{\textsubscript{3}}{{As}}{\textsubscript{3}}},\ }\href
  {https://doi.org/10.1103/PhysRevB.102.064405} {\bibfield  {journal} {\bibinfo
   {journal} {Physical Review B}\ }\textbf {\bibinfo {volume} {102}},\ \bibinfo
  {pages} {064405} (\bibinfo {year} {2020})}\BibitemShut {NoStop}%
\bibitem [{\citenamefont {Higuchi}\ \emph {et~al.}(2016)\citenamefont
  {Higuchi}, \citenamefont {Noshima}, \citenamefont {Shirakawa}, \citenamefont
  {Tsubota},\ and\ \citenamefont {Kitagawa}}]{higuchi2016Optical}%
  \BibitemOpen
  \bibfield  {author} {\bibinfo {author} {\bibfnamefont {S.}~\bibnamefont
  {Higuchi}}, \bibinfo {author} {\bibfnamefont {Y.}~\bibnamefont {Noshima}},
  \bibinfo {author} {\bibfnamefont {N.}~\bibnamefont {Shirakawa}}, \bibinfo
  {author} {\bibfnamefont {M.}~\bibnamefont {Tsubota}},\ and\ \bibinfo {author}
  {\bibfnamefont {J.}~\bibnamefont {Kitagawa}},\ }\bibfield  {title} {\bibinfo
  {title} {Optical, transport and magnetic properties of new compound
  {{CeCd}}{$_{3}$}{{P}}{$_{3}$}},\ }\href
  {https://doi.org/10.1088/2053-1591/3/5/056101} {\bibfield  {journal}
  {\bibinfo  {journal} {Materials Research Express}\ }\textbf {\bibinfo
  {volume} {3}},\ \bibinfo {pages} {056101} (\bibinfo {year}
  {2016})}\BibitemShut {NoStop}%
\bibitem [{\citenamefont {Shin}\ \emph {et~al.}(2020)\citenamefont {Shin},
  \citenamefont {Pomjakushin}, \citenamefont {Keller}, \citenamefont {Rosa},
  \citenamefont {Stuhr}, \citenamefont {Niedermayer}, \citenamefont {Sibille},
  \citenamefont {Toth}, \citenamefont {Kim}, \citenamefont {Jang},
  \citenamefont {Son}, \citenamefont {Lee}, \citenamefont {Shang},
  \citenamefont {Medarde}, \citenamefont {Bauer}, \citenamefont {Kenzelmann},\
  and\ \citenamefont {Park}}]{shin2020Magnetic}%
  \BibitemOpen
  \bibfield  {author} {\bibinfo {author} {\bibfnamefont {S.}~\bibnamefont
  {Shin}}, \bibinfo {author} {\bibfnamefont {V.}~\bibnamefont {Pomjakushin}},
  \bibinfo {author} {\bibfnamefont {L.}~\bibnamefont {Keller}}, \bibinfo
  {author} {\bibfnamefont {P.~F.~S.}\ \bibnamefont {Rosa}}, \bibinfo {author}
  {\bibfnamefont {U.}~\bibnamefont {Stuhr}}, \bibinfo {author} {\bibfnamefont
  {C.}~\bibnamefont {Niedermayer}}, \bibinfo {author} {\bibfnamefont
  {R.}~\bibnamefont {Sibille}}, \bibinfo {author} {\bibfnamefont
  {S.}~\bibnamefont {Toth}}, \bibinfo {author} {\bibfnamefont {J.}~\bibnamefont
  {Kim}}, \bibinfo {author} {\bibfnamefont {H.}~\bibnamefont {Jang}}, \bibinfo
  {author} {\bibfnamefont {S.-K.}\ \bibnamefont {Son}}, \bibinfo {author}
  {\bibfnamefont {H.-O.}\ \bibnamefont {Lee}}, \bibinfo {author} {\bibfnamefont
  {T.}~\bibnamefont {Shang}}, \bibinfo {author} {\bibfnamefont
  {M.}~\bibnamefont {Medarde}}, \bibinfo {author} {\bibfnamefont {E.~D.}\
  \bibnamefont {Bauer}}, \bibinfo {author} {\bibfnamefont {M.}~\bibnamefont
  {Kenzelmann}},\ and\ \bibinfo {author} {\bibfnamefont {T.}~\bibnamefont
  {Park}},\ }\bibfield  {title} {\bibinfo {title} {Magnetic structure and
  crystalline electric field effects in the triangular antiferromagnet
  {{CePtAl}}{\textsubscript{4}}{{Ge}}{\textsubscript{2}}},\ }\href
  {https://doi.org/10.1103/PhysRevB.101.224421} {\bibfield  {journal} {\bibinfo
   {journal} {Physical Review B}\ }\textbf {\bibinfo {volume} {101}},\ \bibinfo
  {pages} {224421} (\bibinfo {year} {2020})}\BibitemShut {NoStop}%
\bibitem [{\citenamefont {Gao}\ \emph {et~al.}(2019)\citenamefont {Gao},
  \citenamefont {Chen}, \citenamefont {Tam}, \citenamefont {Huang},
  \citenamefont {Sasmal}, \citenamefont {Adroja}, \citenamefont {Ye},
  \citenamefont {Cao}, \citenamefont {Sala}, \citenamefont {Stone},
  \citenamefont {Baines}, \citenamefont {Verezhak}, \citenamefont {Hu},
  \citenamefont {Chung}, \citenamefont {Xu}, \citenamefont {Cheong},
  \citenamefont {Nallaiyan}, \citenamefont {Spagna}, \citenamefont {Maple},
  \citenamefont {Nevidomskyy}, \citenamefont {Morosan}, \citenamefont {Chen},\
  and\ \citenamefont {Dai}}]{gao2019Experimental}%
  \BibitemOpen
  \bibfield  {author} {\bibinfo {author} {\bibfnamefont {B.}~\bibnamefont
  {Gao}}, \bibinfo {author} {\bibfnamefont {T.}~\bibnamefont {Chen}}, \bibinfo
  {author} {\bibfnamefont {D.~W.}\ \bibnamefont {Tam}}, \bibinfo {author}
  {\bibfnamefont {C.-L.}\ \bibnamefont {Huang}}, \bibinfo {author}
  {\bibfnamefont {K.}~\bibnamefont {Sasmal}}, \bibinfo {author} {\bibfnamefont
  {D.~T.}\ \bibnamefont {Adroja}}, \bibinfo {author} {\bibfnamefont
  {F.}~\bibnamefont {Ye}}, \bibinfo {author} {\bibfnamefont {H.}~\bibnamefont
  {Cao}}, \bibinfo {author} {\bibfnamefont {G.}~\bibnamefont {Sala}}, \bibinfo
  {author} {\bibfnamefont {M.~B.}\ \bibnamefont {Stone}}, \bibinfo {author}
  {\bibfnamefont {C.}~\bibnamefont {Baines}}, \bibinfo {author} {\bibfnamefont
  {J.~A.~T.}\ \bibnamefont {Verezhak}}, \bibinfo {author} {\bibfnamefont
  {H.}~\bibnamefont {Hu}}, \bibinfo {author} {\bibfnamefont {J.-H.}\
  \bibnamefont {Chung}}, \bibinfo {author} {\bibfnamefont {X.}~\bibnamefont
  {Xu}}, \bibinfo {author} {\bibfnamefont {S.-W.}\ \bibnamefont {Cheong}},
  \bibinfo {author} {\bibfnamefont {M.}~\bibnamefont {Nallaiyan}}, \bibinfo
  {author} {\bibfnamefont {S.}~\bibnamefont {Spagna}}, \bibinfo {author}
  {\bibfnamefont {M.~B.}\ \bibnamefont {Maple}}, \bibinfo {author}
  {\bibfnamefont {A.~H.}\ \bibnamefont {Nevidomskyy}}, \bibinfo {author}
  {\bibfnamefont {E.}~\bibnamefont {Morosan}}, \bibinfo {author} {\bibfnamefont
  {G.}~\bibnamefont {Chen}},\ and\ \bibinfo {author} {\bibfnamefont
  {P.}~\bibnamefont {Dai}},\ }\bibfield  {title} {\bibinfo {title}
  {Experimental signatures of a three-dimensional quantum spin liquid in
  effective spin-1/2
  {{Ce}}{\textsubscript{2}}{{Zr}}{\textsubscript{2}}{{O}}{\textsubscript{7}}
  pyrochlore},\ }\href {https://doi.org/10.1038/s41567-019-0577-6} {\bibfield
  {journal} {\bibinfo  {journal} {Nature Physics}\ }\textbf {\bibinfo {volume}
  {15}},\ \bibinfo {pages} {1052} (\bibinfo {year} {2019})}\BibitemShut
  {NoStop}%
\bibitem [{\citenamefont {Gaudet}\ \emph {et~al.}(2019)\citenamefont {Gaudet},
  \citenamefont {Smith}, \citenamefont {Dudemaine}, \citenamefont {Beare},
  \citenamefont {Buhariwalla}, \citenamefont {Butch}, \citenamefont {Stone},
  \citenamefont {Kolesnikov}, \citenamefont {Xu}, \citenamefont {Yahne},
  \citenamefont {Ross}, \citenamefont {Marjerrison}, \citenamefont {Garrett},
  \citenamefont {Luke}, \citenamefont {Bianchi},\ and\ \citenamefont
  {Gaulin}}]{gaudet2019Quantum}%
  \BibitemOpen
  \bibfield  {author} {\bibinfo {author} {\bibfnamefont {J.}~\bibnamefont
  {Gaudet}}, \bibinfo {author} {\bibfnamefont {E.~M.}\ \bibnamefont {Smith}},
  \bibinfo {author} {\bibfnamefont {J.}~\bibnamefont {Dudemaine}}, \bibinfo
  {author} {\bibfnamefont {J.}~\bibnamefont {Beare}}, \bibinfo {author}
  {\bibfnamefont {C.~R.~C.}\ \bibnamefont {Buhariwalla}}, \bibinfo {author}
  {\bibfnamefont {N.~P.}\ \bibnamefont {Butch}}, \bibinfo {author}
  {\bibfnamefont {M.~B.}\ \bibnamefont {Stone}}, \bibinfo {author}
  {\bibfnamefont {A.~I.}\ \bibnamefont {Kolesnikov}}, \bibinfo {author}
  {\bibfnamefont {G.}~\bibnamefont {Xu}}, \bibinfo {author} {\bibfnamefont
  {D.~R.}\ \bibnamefont {Yahne}}, \bibinfo {author} {\bibfnamefont {K.~A.}\
  \bibnamefont {Ross}}, \bibinfo {author} {\bibfnamefont {C.~A.}\ \bibnamefont
  {Marjerrison}}, \bibinfo {author} {\bibfnamefont {J.~D.}\ \bibnamefont
  {Garrett}}, \bibinfo {author} {\bibfnamefont {G.~M.}\ \bibnamefont {Luke}},
  \bibinfo {author} {\bibfnamefont {A.~D.}\ \bibnamefont {Bianchi}},\ and\
  \bibinfo {author} {\bibfnamefont {B.~D.}\ \bibnamefont {Gaulin}},\ }\bibfield
   {title} {\bibinfo {title} {Quantum spin ice dynamics in the dipole-octupole
  pyrochlore magnet
  {{Ce}}{\textsubscript{2}}{{Zr}}{\textsubscript{2}}{{O}}{\textsubscript{7}}},\
  }\href {https://doi.org/10.1103/PhysRevLett.122.187201} {\bibfield  {journal}
  {\bibinfo  {journal} {Physical Review Letters}\ }\textbf {\bibinfo {volume}
  {122}},\ \bibinfo {pages} {187201} (\bibinfo {year} {2019})}\BibitemShut
  {NoStop}%
\bibitem [{\citenamefont {Por{\'e}e}\ \emph {et~al.}(2022)\citenamefont
  {Por{\'e}e}, \citenamefont {Lhotel}, \citenamefont {Petit}, \citenamefont
  {Krajewska}, \citenamefont {Puphal}, \citenamefont {Clark}, \citenamefont
  {Pomjakushin}, \citenamefont {Walker}, \citenamefont {Gauthier},
  \citenamefont {Gawryluk},\ and\ \citenamefont
  {Sibille}}]{poree2022Crystalfield}%
  \BibitemOpen
  \bibfield  {author} {\bibinfo {author} {\bibfnamefont {V.}~\bibnamefont
  {Por{\'e}e}}, \bibinfo {author} {\bibfnamefont {E.}~\bibnamefont {Lhotel}},
  \bibinfo {author} {\bibfnamefont {S.}~\bibnamefont {Petit}}, \bibinfo
  {author} {\bibfnamefont {A.}~\bibnamefont {Krajewska}}, \bibinfo {author}
  {\bibfnamefont {P.}~\bibnamefont {Puphal}}, \bibinfo {author} {\bibfnamefont
  {A.~H.}\ \bibnamefont {Clark}}, \bibinfo {author} {\bibfnamefont
  {V.}~\bibnamefont {Pomjakushin}}, \bibinfo {author} {\bibfnamefont {H.~C.}\
  \bibnamefont {Walker}}, \bibinfo {author} {\bibfnamefont {N.}~\bibnamefont
  {Gauthier}}, \bibinfo {author} {\bibfnamefont {D.~J.}\ \bibnamefont
  {Gawryluk}},\ and\ \bibinfo {author} {\bibfnamefont {R.}~\bibnamefont
  {Sibille}},\ }\bibfield  {title} {\bibinfo {title} {Crystal-field states and
  defect levels in candidate quantum spin ice
  {{Ce}}{\textsubscript{2}}{{Zr}}{\textsubscript{2}}{{O}}{\textsubscript{7}}},\
  }\href {https://doi.org/10.1103/PhysRevMaterials.6.044406} {\bibfield
  {journal} {\bibinfo  {journal} {Physical Review Materials}\ }\textbf
  {\bibinfo {volume} {6}},\ \bibinfo {pages} {044406} (\bibinfo {year}
  {2022})}\BibitemShut {NoStop}%
\bibitem [{\citenamefont {Sidorov}\ \emph {et~al.}(2003)\citenamefont
  {Sidorov}, \citenamefont {Bauer}, \citenamefont {Frederick}, \citenamefont
  {Jeffries}, \citenamefont {Nakatsuji}, \citenamefont {Moreno}, \citenamefont
  {Thompson}, \citenamefont {Maple},\ and\ \citenamefont
  {Fisk}}]{sidorov2003Magnetic}%
  \BibitemOpen
  \bibfield  {author} {\bibinfo {author} {\bibfnamefont {V.~A.}\ \bibnamefont
  {Sidorov}}, \bibinfo {author} {\bibfnamefont {E.~D.}\ \bibnamefont {Bauer}},
  \bibinfo {author} {\bibfnamefont {N.~A.}\ \bibnamefont {Frederick}}, \bibinfo
  {author} {\bibfnamefont {J.~R.}\ \bibnamefont {Jeffries}}, \bibinfo {author}
  {\bibfnamefont {S.}~\bibnamefont {Nakatsuji}}, \bibinfo {author}
  {\bibfnamefont {N.~O.}\ \bibnamefont {Moreno}}, \bibinfo {author}
  {\bibfnamefont {J.~D.}\ \bibnamefont {Thompson}}, \bibinfo {author}
  {\bibfnamefont {M.~B.}\ \bibnamefont {Maple}},\ and\ \bibinfo {author}
  {\bibfnamefont {Z.}~\bibnamefont {Fisk}},\ }\bibfield  {title} {\bibinfo
  {title} {Magnetic phase diagram of the ferromagnetic {{Kondo-lattice}}
  compound {{CeAgSb}}{$_2$} up to 80 kbar},\ }\href
  {https://doi.org/10.1103/PhysRevB.67.224419} {\bibfield  {journal} {\bibinfo
  {journal} {Physical Review B}\ }\textbf {\bibinfo {volume} {67}},\ \bibinfo
  {pages} {224419} (\bibinfo {year} {2003})}\BibitemShut {NoStop}%
\bibitem [{\citenamefont {Coqblin}(1977)}]{coqblin1977electronic}%
  \BibitemOpen
  \bibfield  {author} {\bibinfo {author} {\bibfnamefont {B.}~\bibnamefont
  {Coqblin}},\ }\bibfield  {title} {\bibinfo {title} {Electronic structure of
  rare-earth metals and {{Alloys}}--the magnetic heavy rare-earths},\
  }\href@noop {} {\bibfield  {journal} {\bibinfo  {journal} {Academic Press
  Inc., New York and London. 1977}\ } (\bibinfo {year} {1977})}\BibitemShut
  {NoStop}%
\bibitem [{\citenamefont {Jobiliong}\ \emph {et~al.}(2005)\citenamefont
  {Jobiliong}, \citenamefont {Brooks}, \citenamefont {Choi}, \citenamefont
  {Lee},\ and\ \citenamefont {Fisk}}]{jobiliong2005Magnetization}%
  \BibitemOpen
  \bibfield  {author} {\bibinfo {author} {\bibfnamefont {E.}~\bibnamefont
  {Jobiliong}}, \bibinfo {author} {\bibfnamefont {J.~S.}\ \bibnamefont
  {Brooks}}, \bibinfo {author} {\bibfnamefont {E.~S.}\ \bibnamefont {Choi}},
  \bibinfo {author} {\bibfnamefont {H.}~\bibnamefont {Lee}},\ and\ \bibinfo
  {author} {\bibfnamefont {Z.}~\bibnamefont {Fisk}},\ }\bibfield  {title}
  {\bibinfo {title} {Magnetization and electrical-transport investigation of
  the dense {{Kondo}} system {{CeAgSb}}{$_{2}$}},\ }\href
  {https://doi.org/10.1103/PhysRevB.72.104428} {\bibfield  {journal} {\bibinfo
  {journal} {Physical Review B}\ }\textbf {\bibinfo {volume} {72}},\ \bibinfo
  {pages} {104428} (\bibinfo {year} {2005})}\BibitemShut {NoStop}%
\bibitem [{\citenamefont {Souza}\ \emph {et~al.}(2016)\citenamefont {Souza},
  \citenamefont {Paupitz}, \citenamefont {Seridonio},\ and\ \citenamefont
  {Lagos}}]{souza2016Specific}%
  \BibitemOpen
  \bibfield  {author} {\bibinfo {author} {\bibfnamefont {M.}~\bibnamefont
  {Souza}}, \bibinfo {author} {\bibfnamefont {R.}~\bibnamefont {Paupitz}},
  \bibinfo {author} {\bibfnamefont {A.}~\bibnamefont {Seridonio}},\ and\
  \bibinfo {author} {\bibfnamefont {R.~E.}\ \bibnamefont {Lagos}},\ }\bibfield
  {title} {\bibinfo {title} {Specific heat anomalies in solids described by a
  multilevel model},\ }\href {https://doi.org/10.1007/s13538-016-0404-9}
  {\bibfield  {journal} {\bibinfo  {journal} {Brazilian Journal of Physics}\
  }\textbf {\bibinfo {volume} {46}},\ \bibinfo {pages} {206} (\bibinfo {year}
  {2016})}\BibitemShut {NoStop}%
\bibitem [{\citenamefont {Fleury}\ and\ \citenamefont
  {Loudon}(1968)}]{FLEURY1968}%
  \BibitemOpen
  \bibfield  {author} {\bibinfo {author} {\bibfnamefont {P.~A.}\ \bibnamefont
  {Fleury}}\ and\ \bibinfo {author} {\bibfnamefont {R.}~\bibnamefont
  {Loudon}},\ }\bibfield  {title} {\bibinfo {title} {{Scattering of Light by
  One- and Two-Magnon Excitations}},\ }\href
  {https://doi.org/10.1103/PhysRev.166.514} {\bibfield  {journal} {\bibinfo
  {journal} {Phys. Rev.}\ }\textbf {\bibinfo {volume} {166}},\ \bibinfo {pages}
  {514} (\bibinfo {year} {1968})}\BibitemShut {NoStop}%
\bibitem [{\citenamefont {Inoue}\ and\ \citenamefont
  {Moriya}(1970)}]{Inoue1970}%
  \BibitemOpen
  \bibfield  {author} {\bibinfo {author} {\bibfnamefont {M.}~\bibnamefont
  {Inoue}}\ and\ \bibinfo {author} {\bibfnamefont {T.}~\bibnamefont {Moriya}},\
  }\bibfield  {title} {\bibinfo {title} {{Raman Scattering by Magnons in Rare
  Earth Metals}},\ }\href {https://doi.org/10.1143/JPSJ.29.117} {\bibfield
  {journal} {\bibinfo  {journal} {Journal of the Physical Society of Japan}\
  }\textbf {\bibinfo {volume} {29}},\ \bibinfo {pages} {117} (\bibinfo {year}
  {1970})}\BibitemShut {NoStop}%
\bibitem [{\citenamefont {Benfatto}\ \emph {et~al.}(2006)\citenamefont
  {Benfatto}, \citenamefont {Silva~Neto}, \citenamefont {Gozar}, \citenamefont
  {Dennis}, \citenamefont {Blumberg}, \citenamefont {Miller}, \citenamefont
  {Komiya},\ and\ \citenamefont {Ando}}]{Benfatto2006}%
  \BibitemOpen
  \bibfield  {author} {\bibinfo {author} {\bibfnamefont {L.}~\bibnamefont
  {Benfatto}}, \bibinfo {author} {\bibfnamefont {M.~B.}\ \bibnamefont
  {Silva~Neto}}, \bibinfo {author} {\bibfnamefont {A.}~\bibnamefont {Gozar}},
  \bibinfo {author} {\bibfnamefont {B.~S.}\ \bibnamefont {Dennis}}, \bibinfo
  {author} {\bibfnamefont {G.}~\bibnamefont {Blumberg}}, \bibinfo {author}
  {\bibfnamefont {L.~L.}\ \bibnamefont {Miller}}, \bibinfo {author}
  {\bibfnamefont {S.}~\bibnamefont {Komiya}},\ and\ \bibinfo {author}
  {\bibfnamefont {Y.}~\bibnamefont {Ando}},\ }\bibfield  {title} {\bibinfo
  {title} {{Field dependence of the magnetic spectrum in anisotropic and
  Dzyaloshinskii-Moriya antiferromagnets. II. Raman spectroscopy}},\ }\href
  {https://doi.org/10.1103/PhysRevB.74.024416} {\bibfield  {journal} {\bibinfo
  {journal} {Phys. Rev. B}\ }\textbf {\bibinfo {volume} {74}},\ \bibinfo
  {pages} {024416} (\bibinfo {year} {2006})}\BibitemShut {NoStop}%
\bibitem [{\citenamefont {Andersen}\ and\ \citenamefont
  {Smith}(1979)}]{andersen1979Electronmagnon}%
  \BibitemOpen
  \bibfield  {author} {\bibinfo {author} {\bibfnamefont {N.~H.}\ \bibnamefont
  {Andersen}}\ and\ \bibinfo {author} {\bibfnamefont {H.}~\bibnamefont
  {Smith}},\ }\bibfield  {title} {\bibinfo {title} {Electron-magnon interaction
  and the electrical resistivity of {{Tb}}},\ }\href
  {https://doi.org/10.1103/PhysRevB.19.384} {\bibfield  {journal} {\bibinfo
  {journal} {Physical Review B}\ }\textbf {\bibinfo {volume} {19}},\ \bibinfo
  {pages} {384} (\bibinfo {year} {1979})}\BibitemShut {NoStop}%
\bibitem [{\citenamefont {Kadowaki}\ and\ \citenamefont
  {Woods}(1986)}]{kadowaki1986Universal}%
  \BibitemOpen
  \bibfield  {author} {\bibinfo {author} {\bibfnamefont {K.}~\bibnamefont
  {Kadowaki}}\ and\ \bibinfo {author} {\bibfnamefont {S.}~\bibnamefont
  {Woods}},\ }\bibfield  {title} {\bibinfo {title} {Universal relationship of
  the resistivity and specific heat in heavy-{{Fermion}} compounds},\ }\href
  {https://doi.org/10.1016/0038-1098(86)90785-4} {\bibfield  {journal}
  {\bibinfo  {journal} {Solid State Communications}\ }\textbf {\bibinfo
  {volume} {58}},\ \bibinfo {pages} {507} (\bibinfo {year} {1986})}\BibitemShut
  {NoStop}%
\bibitem [{\citenamefont {Jacko}\ \emph {et~al.}(2009)\citenamefont {Jacko},
  \citenamefont {Fj{\ae}restad},\ and\ \citenamefont
  {Powell}}]{jacko2009Unified}%
  \BibitemOpen
  \bibfield  {author} {\bibinfo {author} {\bibfnamefont {A.~C.}\ \bibnamefont
  {Jacko}}, \bibinfo {author} {\bibfnamefont {J.~O.}\ \bibnamefont
  {Fj{\ae}restad}},\ and\ \bibinfo {author} {\bibfnamefont {B.~J.}\
  \bibnamefont {Powell}},\ }\bibfield  {title} {\bibinfo {title} {A unified
  explanation of the {{Kadowaki}}--{{Woods}} ratio in strongly correlated
  metals},\ }\href {https://doi.org/10.1038/nphys1249} {\bibfield  {journal}
  {\bibinfo  {journal} {Nature Physics}\ }\textbf {\bibinfo {volume} {5}},\
  \bibinfo {pages} {422} (\bibinfo {year} {2009})}\BibitemShut {NoStop}%
\bibitem [{\citenamefont {Stevens}(1952)}]{stevens1952Matrix}%
  \BibitemOpen
  \bibfield  {author} {\bibinfo {author} {\bibfnamefont {K.~W.~H.}\
  \bibnamefont {Stevens}},\ }\bibfield  {title} {\bibinfo {title} {Matrix
  elements and operator equivalents connected with the magnetic properties of
  rare earth ions},\ }\href {https://doi.org/10.1088/0370-1298/65/3/308}
  {\bibfield  {journal} {\bibinfo  {journal} {Proceedings of the Physical
  Society. Section A}\ }\textbf {\bibinfo {volume} {65}},\ \bibinfo {pages}
  {209} (\bibinfo {year} {1952})}\BibitemShut {NoStop}%
\bibitem [{\citenamefont {Hutchings}(1964)}]{hutchings1964Pointcharge}%
  \BibitemOpen
  \bibfield  {author} {\bibinfo {author} {\bibfnamefont {M.}~\bibnamefont
  {Hutchings}},\ }\bibfield  {title} {\bibinfo {title} {Point-charge
  calculations of energy levels of magnetic ions in crystalline electric
  fields},\ }\href {https://doi.org/10.1016/S0081-1947(08)60517-2} {\bibfield
  {journal} {\bibinfo  {journal} {Solid State Physics}\ }\textbf {\bibinfo
  {volume} {16}},\ \bibinfo {pages} {227} (\bibinfo {year} {1964})}\BibitemShut
  {NoStop}%
\bibitem [{\citenamefont {Bowden}\ \emph {et~al.}(1971)\citenamefont {Bowden},
  \citenamefont {Bunbury},\ and\ \citenamefont
  {McCausland}}]{bowden1971Crystal}%
  \BibitemOpen
  \bibfield  {author} {\bibinfo {author} {\bibfnamefont {G.~J.}\ \bibnamefont
  {Bowden}}, \bibinfo {author} {\bibfnamefont {D.~S.~P.}\ \bibnamefont
  {Bunbury}},\ and\ \bibinfo {author} {\bibfnamefont {M.~A.~H.}\ \bibnamefont
  {McCausland}},\ }\bibfield  {title} {\bibinfo {title} {Crystal fields and
  magnetic anisotropy in the molecular field approximation. {{I}}. {{General}}
  considerations},\ }\href {https://doi.org/10.1088/0022-3719/4/13/035}
  {\bibfield  {journal} {\bibinfo  {journal} {Journal of Physics C: Solid State
  Physics}\ }\textbf {\bibinfo {volume} {4}},\ \bibinfo {pages} {1840}
  (\bibinfo {year} {1971})}\BibitemShut {NoStop}%
\bibitem [{\citenamefont {Wang}(1971)}]{wang1971Crystalfield}%
  \BibitemOpen
  \bibfield  {author} {\bibinfo {author} {\bibfnamefont {Y.-L.}\ \bibnamefont
  {Wang}},\ }\bibfield  {title} {\bibinfo {title} {Crystal-field effects of
  paramagnetic {{Curie}} temperature},\ }\href
  {https://doi.org/10.1016/0375-9601(71)90750-X} {\bibfield  {journal}
  {\bibinfo  {journal} {Physics Letters A}\ }\textbf {\bibinfo {volume} {35}},\
  \bibinfo {pages} {383} (\bibinfo {year} {1971})}\BibitemShut {NoStop}%
\bibitem [{\citenamefont {Kabeya}\ \emph {et~al.}(2022)\citenamefont {Kabeya},
  \citenamefont {Takahara}, \citenamefont {Arisumi}, \citenamefont {Kimura},
  \citenamefont {Araki}, \citenamefont {Katoh},\ and\ \citenamefont
  {Ochiai}}]{kabeya2022Eigenstate}%
  \BibitemOpen
  \bibfield  {author} {\bibinfo {author} {\bibfnamefont {N.}~\bibnamefont
  {Kabeya}}, \bibinfo {author} {\bibfnamefont {S.}~\bibnamefont {Takahara}},
  \bibinfo {author} {\bibfnamefont {T.}~\bibnamefont {Arisumi}}, \bibinfo
  {author} {\bibfnamefont {S.}~\bibnamefont {Kimura}}, \bibinfo {author}
  {\bibfnamefont {K.}~\bibnamefont {Araki}}, \bibinfo {author} {\bibfnamefont
  {K.}~\bibnamefont {Katoh}},\ and\ \bibinfo {author} {\bibfnamefont
  {A.}~\bibnamefont {Ochiai}},\ }\bibfield  {title} {\bibinfo {title}
  {Eigenstate analysis of the crystal electric field at low-symmetry sites:
  {{Application}} for an orthogonal site in the tetragonal crystal
  {{Ce}}{$_{2}$}{{Pd}}{$_{2}$}{{Pb}}},\ }\href
  {https://doi.org/10.1103/PhysRevB.105.014419} {\bibfield  {journal} {\bibinfo
   {journal} {Physical Review B}\ }\textbf {\bibinfo {volume} {105}},\ \bibinfo
  {pages} {014419} (\bibinfo {year} {2022})}\BibitemShut {NoStop}%
\bibitem [{\citenamefont {Thalmeier}\ and\ \citenamefont
  {Fulde}(1982)}]{Thalmeier1982}%
  \BibitemOpen
  \bibfield  {author} {\bibinfo {author} {\bibfnamefont {P.}~\bibnamefont
  {Thalmeier}}\ and\ \bibinfo {author} {\bibfnamefont {P.}~\bibnamefont
  {Fulde}},\ }\bibfield  {title} {\bibinfo {title} {{Bound State between a
  Crystal-Field Excitation and a Phonon in Ce${\mathrm{Al}}_{2}$}},\ }\href
  {https://doi.org/10.1103/PhysRevLett.49.1588} {\bibfield  {journal} {\bibinfo
   {journal} {Phys. Rev. Lett.}\ }\textbf {\bibinfo {volume} {49}},\ \bibinfo
  {pages} {1588} (\bibinfo {year} {1982})}\BibitemShut {NoStop}%
\bibitem [{\citenamefont {Thalmeier}(1984)}]{Thalmeier1984}%
  \BibitemOpen
  \bibfield  {author} {\bibinfo {author} {\bibfnamefont {P.}~\bibnamefont
  {Thalmeier}},\ }\bibfield  {title} {\bibinfo {title} {{Theory of the bound
  state between phonons and a CEF excitation in Ce${\mathrm{Al}}_{2}$}},\
  }\href {https://doi.org/10.1088/0022-3719/17/23/015} {\bibfield  {journal}
  {\bibinfo  {journal} {Journal of Physics C: Solid State Physics}\ }\textbf
  {\bibinfo {volume} {17}},\ \bibinfo {pages} {4153} (\bibinfo {year}
  {1984})}\BibitemShut {NoStop}%
\end{thebibliography}

%

\end{document}